\documentclass[aps,prd,twocolumn,preprintnumbers,groupedaddress,superscriptaddress,floatfix,tightenlines,reprint,nofootinbib,longbibliography]{revtex4-1}
\usepackage{subfigure,amsmath,amsthm,amsfonts,amssymb,hyperref,graphicx,bbm,bm, enumitem,slashed}

\usepackage{color}
\def\RR{\mathbb R}

\newcommand{\ZZ}{\mathbb{Z}}

\begin{document}
\title{Modified Villain formulation of abelian Chern-Simons theory}

%\author{Theodore Jacobson$^1$\footnote{jaco2585@umn.edu} \ and Tin Sulejmanpasic$^2$\footnote{tin.sulejmanpasic@durham.ac.uk}\\ \\ 
%{\it $^1$ School of Physics and Astronomy, University of Minnesota,}\\
%{\it Minneapolis, MN 55455 USA} \\
%{\it $^2$ Department of Mathematical Sciences, Durham University,} \\ 
%{\it DH1 3LE Durham, UK }
%}

\author{Theodore Jacobson}
\email{jaco2585@umn.edu}
\affiliation{School of Physics and Astronomy, University of Minnesota, Minneapolis, MN 55455 USA}

\author{Tin Sulejmanpasic}
\email{tin.sulejmanpasic@durham.ac.uk}
\affiliation{Department of Mathematical Sciences, Durham University, DH1 3LE Durham, UK}

\begin{abstract}
We formulate $U(1)_k$ Chern-Simons theory on a Euclidean spacetime lattice using the modified Villain approach. Various familiar aspects of continuum Chern-Simons theory such as level quantization, framing, the discrete 1-form symmetry and its 't Hooft anomaly, as well as the electric charge of monopole operators are manifest in our construction. The key technical ingredient is the cup product and its higher generalizations on the (hyper-)cubic lattice, which recently appeared in the literature. All unframed Wilson loops are projected out by a peculiar subsystem symmetry, leaving topological, ribbon-like Wilson loops which have the correct correlation functions and topological spins expected from the continuum theory. 
%Like most other discretizations of abelian Chern-Simons theory, our construction has a peculiar symmetry under which all naive (unframed) Wilson loops are charged. As a result, unframed Wilson loops are identically zero in this theory. However, we identify ribbon-like Wilson loops which are neutral under this symmetry. Such Wilson loops are topological and have the correct correlation functions and topological spin expected from the continuum theory. 
Our action can be obtained from a new definition of the theta term in four dimensions which improves upon previous constructions within the modified Villain approach. This bulk action coupled to background fields for the 1-form symmetry is given by the Pontryagin square, which provides anomaly inflow directly on the lattice. 
\end{abstract}

\maketitle

%%%%%%%%%%%%%
\section{Introduction}
%%%%%%%%%%%%%

Despite its ubiquity and apparent simplicity in the continuum, it is not obvious that abelian Chern-Simons (CS) theory admits a lattice regularization. Indeed, there are claims in the literature that the most basic $U(1)$ CS theory, with continuum action
\begin{equation}
S = \frac{ik}{4\pi} \int a \wedge da\,,
\end{equation}
cannot be formulated in a local way on the lattice~\cite{Kapustin:2014zva,Chen:2019mjw,Radicevic:2021fnm}, with the culprit often identified as the framing anomaly~\cite{Polyakov:1988md,Witten:1988hf} or chiral central charge~\cite{Kitaev:2006lla}. A direct consequence of the framing anomaly is that Wilson loops require point-splitting regularization to be well-defined. The physical operators in continuum CS theory are therefore ribbons, or framed Wilson loops, rather than standard line operators. One might hope that a fully regularized lattice formulation of CS theory would help illuminate precisely such subtleties of the continuum theory which make it difficult to discretize in the first place. Aside from providing a setting to study aspects of CS theory on its own, such a lattice description could be useful in demonstrating exact boson/fermion dualities, constructing non-invertible defects in four-dimensional theories, and has some parallels to the problem of putting chiral fermions on the lattice.  

In fact there is a long history of attempts to discretize CS theory on Euclidean spacetime lattices~\cite{Frohlich:1988qh,Kavalov:1989kg,Diamantini:1993iu, DeMarco:2019pqv,Zhang:2021bqo} as well as in the Hamiltonian framework where time is kept continuous~\cite{Luscher:1989kk,Muller:1990xd,Eliezer:1991qh,Eliezer:1992sq,Sun:2015hla}. However, perhaps surprisingly, global aspects have been all but ignored in the literature. The main goal of this paper is to provide a discretization of $U(1)_k$ CS theory that correctly captures its global features such as its symmetries, level quantization, framing, and the role of monopoles directly on the lattice. Our construction is based on the modified Villain approach~\cite{Villain:1974ir,Gross:1990ub,Sulejmanpasic:2019ytl,Gorantla:2021svj} which naturally endows certain lattice theories with features (such as symmetries, dualities, and anomalies) of their continuum limits (see also \cite{Goschl:2018uma, Anosova:2019quw, Sulejmanpasic:2020lyq, Gattringer:2019yof, Sulejmanpasic:2020ubo, Anosova:2021akr,Choi:2021kmx,Choi:2022zal,Anosova:2022cjm,Anosova:2022yqx,Hirtler:2022ycl,Fazza:2022fss} for related works).  

Chern-Simons theory has no interesting local dynamics. It is therefore crucial for any formulation of CS theory to incorporate its global aspects, which are all that remain. In the present abelian context the fact that we consider a compact (i.e. $U(1)$ rather than $\RR$) gauge group means that one can have quantized magnetic fluxes, 
\begin{equation}\label{eq:quantized_flux} 
\int_{\Sigma} da \in 2\pi \ZZ\,,
\end{equation}
where $a$ is the $U(1)$ gauge field and $\Sigma$ is a closed surface. If the surface is contractible, the above equation indicates the presence of a monopole somewhere in its interior. In the continuum, it is well-known that such monopole configurations are not gauge invariant in the presence of a CS term~\cite{PhysRevD.34.3851,AFFLECK1989575}. This might appear to pose a problem for formulating CS theory in a fully gauge-invariant way on the lattice, as generic discretizations of $U(1)$ gauge theory contain dynamical lattice-scale monopoles. 

The modified Villain approach circumvents this issue by offering complete control over monopoles. In the conventional Villain or `periodic Gaussian' formulation quantized magnetic flux is encoded in discrete plaquette variables $n\in\ZZ$ in addition to the familiar algebra-valued gauge fields $a\in\RR$ living on links~\cite{Villain:1974ir}. The plaquette variable $n$ can be interpreted as a discrete gauge field for the $\ZZ$ 1-form symmetry of the pure, noncompact $\RR$
gauge theory which acts by $a \to a + 2\pi$. Gauging these discrete shifts is equivalent to studying compact $U(1) = \RR/2\pi\ZZ$ gauge theory. In the modified Villain formulation, monopoles are consequently eliminated from the theory by introducing a Lagrange multiplier which constrains the discrete gauge field to be flat~\cite{Sulejmanpasic:2019ytl}. This modification allows one to establish various dualities directly on the lattice, where depending on the context the Lagrange multiplier assumes the role of a T-dual scalar, dual photon, or magnetic gauge field. This approach has also found applications in elucidating the behavior of fracton models~\cite{Gorantla:2021svj,Yoneda:2022qpj} and has been recently generalized to the Hamiltonian formulation~\cite{Fazza:2022fss,Cheng:2022sgb}. 

\begin{figure}[h] 
\centering
\includegraphics[width=0.4\textwidth]{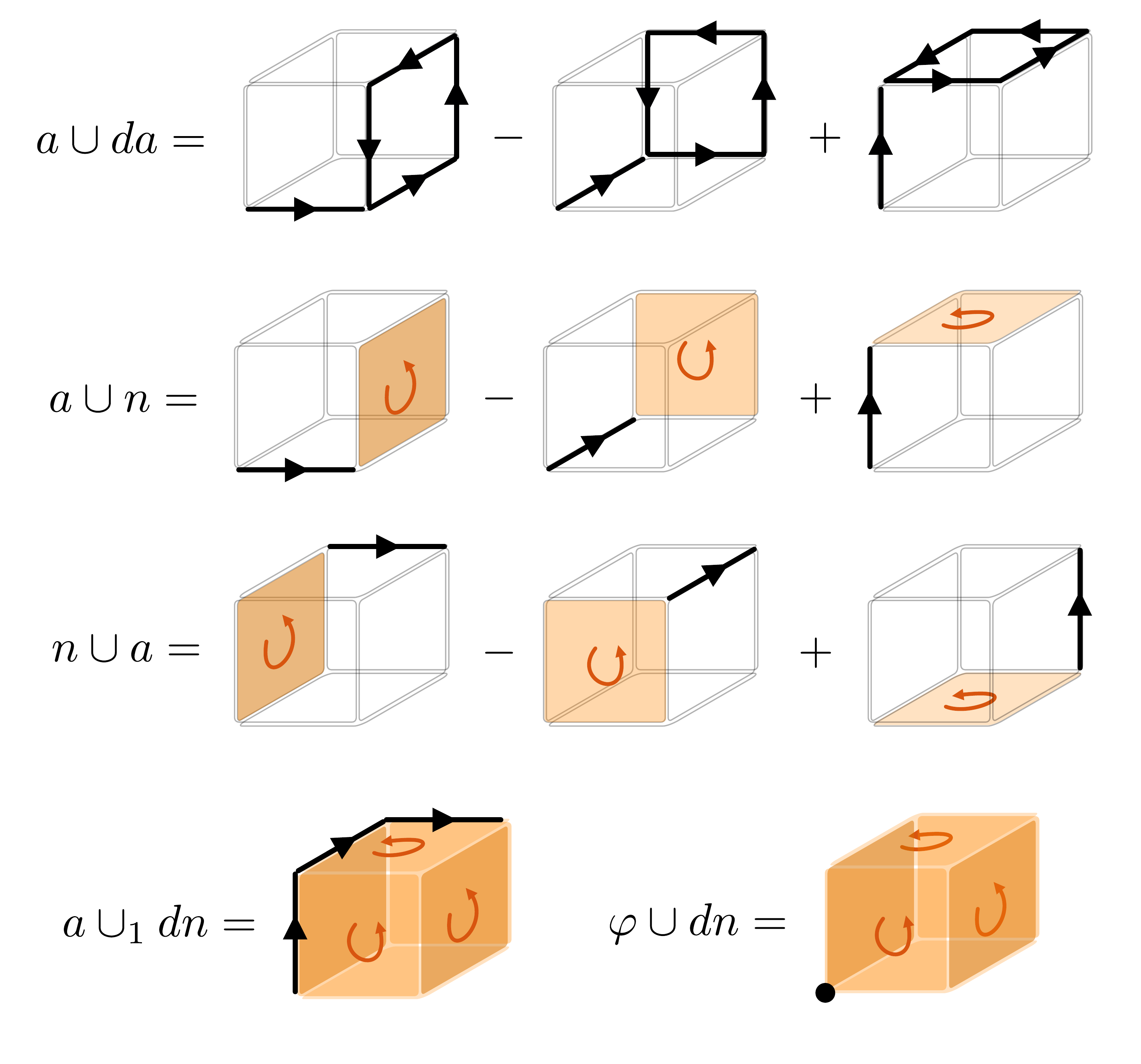}
\caption{Cup products used in the CS action \eqref{eq:action}. The black lines represent the gauge field $a$, the orange plaquettes represent the discrete magnetic flux variable $n$, and the black dot represents the Lagrange multiplier $\varphi$. } 
\label{fig:action_ingredients}
\end{figure}

Our lattice action can be written compactly in terms of (higher) cup products as follows:
\begin{subequations}\label{eq:action_both}
\begin{equation}
\begin{split} 
\label{eq:action_0} 
S(a,n,\varphi) = \sum_c \frac{ik}{4\pi} \left[ a\cup da - 2\pi (a\cup n + n \cup a) \right] \\
-\frac{ik}{2}a \cup_1 dn + i \varphi \cup dn\,,
\end{split}
\end{equation}
where the sum is over all cubes of the lattice, and $\varphi$ is the aforementioned Lagrange multiplier which removes monopoles. We give explicit expressions for the higher cup products in App.~\ref{sec:products}---graphical representations of each of the terms appearing above are shown in Fig.~\ref{fig:action_ingredients}. In a more conventional lattice gauge theory notation, our action reads
\begin{align} \label{eq:action_1} 
&S(a,n,\varphi) =\\
&\frac{ik}{4\pi} \sum_{\substack{x,\mu \\ \nu<\rho}} \epsilon_{\mu\nu\rho}\Big[  a_{x,\mu} \left( da- 2\pi \,  n\right)_{x+\hat\mu,\nu\rho} - 2\pi\,  n_{x,\nu\rho}\, a_{x+\hat\nu + \hat\rho,\mu} \Big] \nonumber \\
&+i \sum_x \left[\varphi_x-\frac{k}{2}(a_{x,3}+a_{x+\hat{3},2} + a_{x+\hat{3}+\hat{2},1})\right] (dn)_{x,123} \,, \nonumber
\end{align}
\end{subequations}
where the sum is over all sites $x$ on the lattice, $\mu,\nu, \rho \in \{ 1,2,3\}$ and $\hat\mu$ denotes a unit vector in the $\mu$ direction---cells are labelled by a `root' site and the directions in which the cell extends.  Our notation is explained in more detail below. It should be clear from this form of the action that the $\cup$ and $\cup_1$ products explicitly break the discrete rotational invariance of the lattice. 

The action \eqref{eq:action_both} turns out to have a peculiar staggered symmetry\footnote{The symmetry is akin to a subsystem symmetry, where it transforms fields on links related by a diagonal lattice translation (see Fig.~\ref{fig:staggeredsymmetry} and discussion around it). The precise form of the symmetry depends on the definition of the cup product.} commonly associated with Chern-Simons discretizations. This staggered symmetry causes extra zero modes to appear in the Gaussian operator. This was shown to be generic for any local, gauge-invariant, parity-odd Euclidean lattice action~\cite{Berruto:2000dp}, and has been likened to the well-known fermion doubling problem associated with putting chiral fermions on the lattice~\cite{Nielsen:1981hk}.\footnote{The zero modes may be lifted by including additional terms in such a way that the action is invariant under a modified parity transformation~\cite{Bietenholz:2002mt,Bietenholz:2003vw}, analogous to the Ginsparg-Wilson approach to chiral fermions on the lattice~\cite{Ginsparg:1981bj,Luscher:1998pqa,Neuberger:1997fp}. However, it is not clear if the resulting theory shares the desired topological properties of continuum CS theory.}

This symmetry has important consequences. It implies that the non-trivial gauge-invariant observables in our theory are in fact \emph{framed} Wilson lines, or ribbons. These ribbons are topological and have fully computable correlation functions. An example is shown below in Fig.~\ref{fig:topologicalspin}. The curve $\tilde C_{\text{twist}}$ shown there is twisted in a precise sense: the corresponding ribbon has non-trivial self-linking computed with our fixed choice of framing. 

\begin{figure}[h] 
\centering
\includegraphics[width=0.36\textwidth]{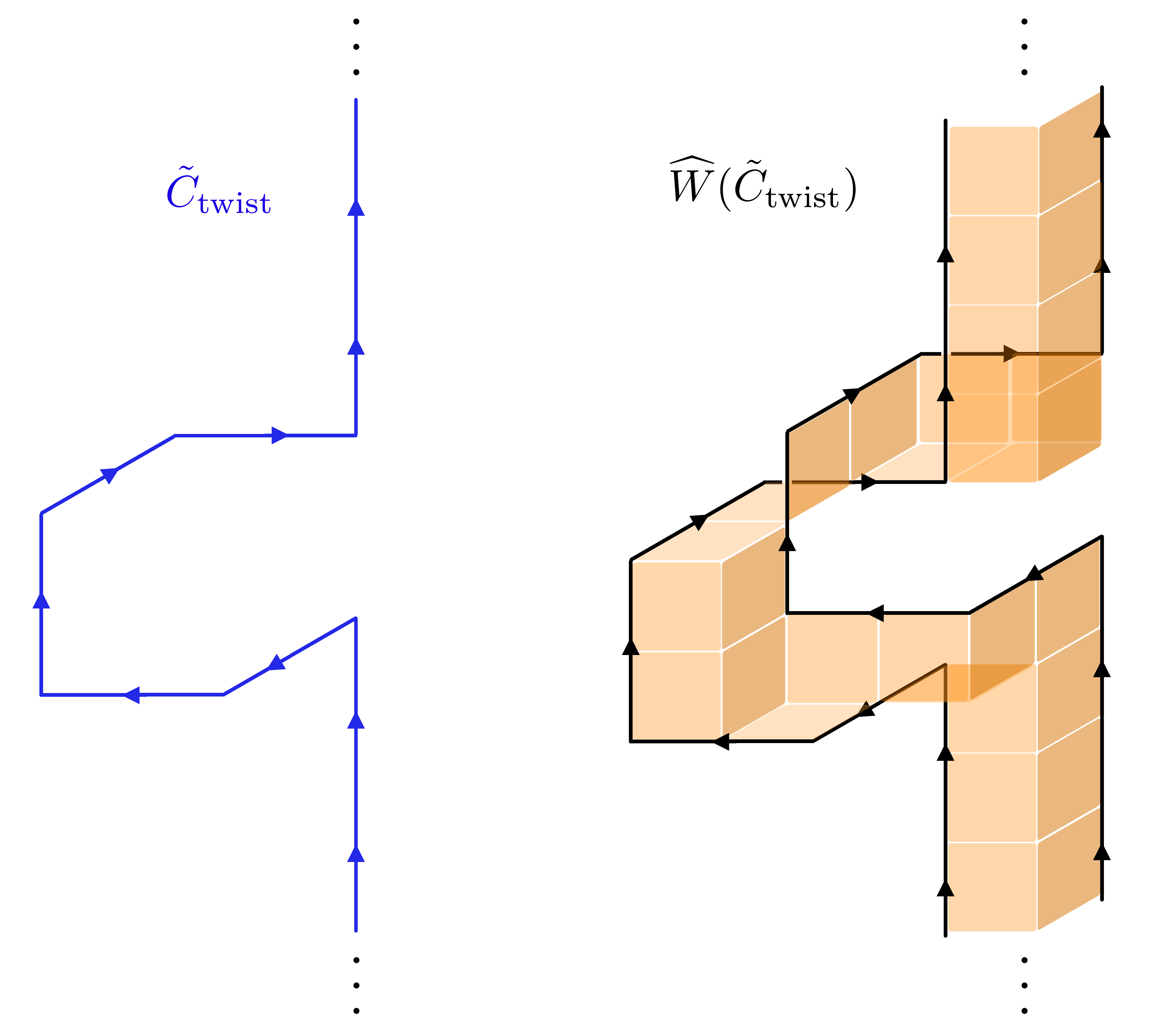}
\caption{A framed Wilson loop defined on a curve $\tilde C_{\text{twist}}$ with a non-vanishing self-linking number. The black lines denote `ordinary' charge-$1/2$ Wilson lines of the dynamical gauge field $a$, and the orange plaquettes indicate the support of surface operators built from the discrete gauge field $n$. } 
\label{fig:topologicalspin}
\end{figure}

Much of the foundational literature on lattice CS theory viewed the zero modes associated with the aforementioned staggered symmetry as detrimental. This is not without reason---they make canonical quantization more subtle. But they are also avoidable in the Hamiltonian formulation---in a set of papers~\cite{Eliezer:1991qh,Eliezer:1992sq} Eliezer and Semenoff were able to construct and solve a gauge-invariant local lattice Hamiltonian free of extra zero modes. This was possible because they included couplings between adjacent \emph{parallel} link variables, which disappear as one takes the lattice spacing to zero. Their solution matches much of the physics of continuum CS theory, but still suffers from ambiguities related to the self-intersection of Wilson lines. Although for reasons of brevity we will not discuss it here, one can show that these ambiguities can be resolved by discretizing the time direction at the cost of reintroducing the zero modes. Finally, one can easily put ``doubled" CS theories on the lattice without encountering extra zero modes~\cite{Kantor:1991ty,Adams:1997eb,Adams:1996yf,Olesen:2015baa,Chen:2019mjw,Banks:2021igc}.

We reiterate that in stark contrast to the older literature, our point of view is that the presence of the zero modes and the associated staggered symmetry on a space-time lattice is not a problem. In fact, the staggered symmetry projects out all of the naive Wilson loops! This as a blessing, rather than a curse, since it directly reflects the fact that the continuum Chern-Simons theory has a framing anomaly which forces one to pick a framing for every loop. In other words, observables in Chern-Simons theory are not loops, but strips. In fact we show that the correct topological observable on the lattice is a Wilson strip, which can be viewed as two parallel charge-$1/2$ Wilson loops connected by a surface.\footnote{A fractional Wilson loop is not well-defined without a surface.} As we will see, all key ingredients of this construction agree with the expectations from the continuum.

%    On a spacetime lattice, the additional zero modes present in naive discretizations pose no obstacle to computing Euclidean correlation functions of gauge-invariant observables---namely framed Wilson loops. In fact, as we argue below, in an Euclidean formulation these zero modes should be viewed as a blessing rather than a curse, as they perform the function of projecting out all Wilson loops which are not framed. Therefore, in this paper we are content presenting a construction with lattice zero modes, as we are still able to discuss the relevant features of the model and unambiguously compute correlators of physical operators. 

Before moving on, let us make some more detailed remarks on related recent works. The older literature did not incorporate the crucial global aspects of CS theory, with the exception of Ref.~\cite{Eliezer:1992sq} which took into account large gauge transformations by hand to canonically quantize the theory on the torus. More recently, Refs.~\cite{DeMarco:2019pqv,Chen:2019mjw} gave lattice constructions of abelian CS theories with multiple $U(1)$ factors, taking into account the compactness of the gauge group. See also Ref.~\cite{Kobayashi:2021jsc} for a recent application of the Villain formulation to the study of anomalies in 2+1D topological phases.

Reference~\cite{DeMarco:2019pqv} presented a discretization of $U(1)^{K}$ CS theory on a triangulation and showed that their action preserves the 1-form global symmetries of the continuum theory. The dynamical variables in their construction are simply the real-valued gauge fields $a_I$, $I = 1 , \ldots, K$ which live on each link. The quantized magnetic flux is then, schematically, 
\begin{equation}
2\pi\int_{\Sigma} \lfloor da_I \rceil \in 2\pi \ZZ,
\end{equation}  
where $\lfloor x \rceil$ denotes the integer nearest to $x$ and $\lfloor da_I \rceil$ represents the quantized magnetic flux through a plaquette. The lattice action of Ref.~\cite{DeMarco:2019pqv} is a non-continuous function of the real-valued variables $a_I$ and is invariant under large gauge transformations $a_I \to a_I +2\pi m_I$ with $m_I \in \ZZ$. However, Ref.~\cite{DeMarco:2019pqv} must include a Maxwell term with a large coefficient to suppress monopole configurations where $d\lfloor da_I \rceil \not =0$. For any nonzero value of the gauge coupling, monopoles exist and spoil ordinary 0-form gauge invariance. The lack of gauge redundancy is pointed out by the authors as a welcome feature of their model, as it allows for a tensor product Hilbert space. In this paper we take invariance under ordinary gauge transformations to be a necessary ingredient.   

Reference~\cite{Chen:2019mjw} employed the Villain approach to construct doubled CS theory (with both compact and non-compact gauge groups) on both cubic and triangulated lattices. 
%To maintain gauge invariance, the integer-valued plaquette variables were constrained \emph{by hand} to be flat. This obscures the role of monopoles and the (un)faithfulness of its global symmetries. Nevertheless, 
In particular, Ref.~\cite{Chen:2019mjw} contains a comprehensive analysis of the doubled CS theory with gauge group $U(1)^2$ and $K$-matrix $\begin{pmatrix} 0 & n \\ n & k \end{pmatrix}$ with $n \in\ZZ$ and $k \in 2\ZZ$, including detailed computations of the partition function and correlation functions on spacetimes with torsion, a reconstruction of the Hilbert space from lattices with boundary, and a method to reproduce the continuum path integral using a correspondence between the Villain formulation on a triangulation and Deligne-Beilinson cohomology.

The remainder of the paper is structured as follows. In Sec.~\ref{sec:action} we briefly review our conventions for cochain (form) notation on the cubic lattice, and present our lattice action. We show how level quantization and the electric charge of monopoles arise from demanding full $U(1)$ gauge invariance. In Sec.~\ref{sec:symmetries} we discuss the symmetries of the theory, which include the $\ZZ_k$ 1-form symmetry and an exotic `staggered' symmetry which projects out ordinary Wilson loops. In Sec.~\ref{sec:backgroundfields} we describe the correspondence between topological, framed Wilson loops (or ribbon operators) and background fields for the 1-form symmetry. We compute the 't Hooft anomaly for the 1-form symmetry and use it to identify twisted Wilson loops. Section~\ref{sec:thetaterm} is dedicated to a novel definition of the theta term on the lattice in four dimensions. When $\theta = 2\pi k$ with $k$ even, we recover our 3d CS theory on the boundary of a 4d lattice. Coupling the bulk to background fields for the 1-form symmetry leads to an anomaly inflow action based on the Pontryagin square. Explicit formulas for the cup products and their higher generalizations, as well as a discussion of the Pontryagin square, are collected in App.~\ref{sec:products} and~\ref{sec:pontryagin}.

%Hamiltonian Maxwell-CS \cite{Luscher:1989kk,Muller:1990xd}
%Euclidean Maxwell-CS \cite{Diamantini:1993iu}
%
%Higher cup products on the cubic lattice \cite{Chen:2021ppt}
%
%First known formula \cite{Frohlich:1988qh} Euclidean, Maxwell-CS. 
%
%Early work, Hamiltonian formulation, bosonization \cite{Luscher:1989kk} already noted the extra zero modes. Hence Maxwell-CS. Also Hamiltonian, anyons, Maxwell-CS \cite{Muller:1990xd}. Euclidean Maxwell-CS \cite{Diamantini:1993iu}, actually did Villain-type construction. 
%
%Solved doubling on Hamiltonian \cite{Eliezer:1991qh,Eliezer:1992sq}. Solved Hamiltonian but there is ambiguity having to do with self-linking. This is resolved in the Euclidean construction, but one introduces zero modes. 
%
%Doubling problem investigated \cite{Berruto:2000dp}, analysis is for Euclidean action not Hamiltonian. Get out of doubling problem by weird parity \cite{Bietenholz:2002mt,Bietenholz:2003vw}
%
%CS theory on a triangulation, Euclidean, no global structure \cite{Kavalov:1989kg}
%
%Hamiltonian on graph, \cite{Sun:2015hla}. 
%
%Higher cup products, Math: \cite{Tata:2020qca} Physics applications: \cite{Chen:2018nog, Chen:2019wlx, Chen:2021ppt} 
%
%Doubled CS theory \cite{Chen:2019mjw} Euclidean, \cite{Olesen:2015baa} lattice Maxwell-CS. \cite{Adams:1997eb,Adams:1996yf} doubled, simplicial and cubic. \cite{Kantor:1991ty} 

%%%%%%%%%%%%%
\section{The modified Villain action}
%%%%%%%%%%%%%
\label{sec:action} 

%%%%%%%%%%%%%
\subsection{Lattice preliminaries}
%%%%%%%%%%%%%

Throughout the paper we use the language of differential forms or cochains on the cubic lattice.\footnote{See e.g. App. A of Ref.~\cite{Sulejmanpasic:2019ytl} for more details regarding differential forms (i.e. cochains) on hypercubic lattices.} We consider three- and four-dimensional periodic lattices (denoted generically by $M$) with lattice spacing set to one. Fields that live on sites (denoted $s$ or $x$), links ($\ell$), plaquettes ($p$), cubes ($c$), and hypercubes ($h$) of the lattice are referred to as 0-, 1-, 2-, 3-, and 4-cochains. In addition, fields can take real or integer values, or can be finite spins taking values only from say $0,1,\dots q-1$ for some integer $q$. These are then naturally associated with abelian groups $\RR$, $\ZZ$ and $\ZZ_q$ (we use additive notation for all group operations). Therefore a field living on a $p$-cell which takes real, integer and integer mod $q$ values are referred to as belonging to the set of $m$-cochains $C^m(M,\RR), C^m(M,\ZZ)$  and $C^m(M,\ZZ_q)$ respectively. Further, there is a natural exterior derivative $d$ of these fields which maps a field on a $m$-cell to a field on a $m+1$ cell (i.e. a $m$-cochain to a $m+1$-cochain). 

If the exterior derivative of an $m$-cochain is zero, then it is called closed while if a $m$-cochain is the exterior derivative of a $(m-1)$-cochain it is called exact in analogy with differential forms. The set of closed $m$-cochains valued in an abelian group $G$, which are called $m$-cocycles, is denoted by $Z^m(M,G)$, while the set of exact $m$-cochains (or $m$-coboundaries) is denoted by $B^m(M,G)$. The $m$-th cohomology class $H^m(M,G)$ is the set of $m$-cocycles which are not coboundaries\footnote{Or in other words, $H^m(M,G)$ is the set of closed $m$-forms valued in $G$ which are not exact.}  $H^m(M,G) = Z^m(M,G)/B^m(M,G)$. We only consider $G = \RR,\ZZ, \ZZ_q$ in this paper. To reduce clutter we will not indicate the degree of a given cochain unless necessary.  

It is often useful to view a given cochain valued in a group $G$ as being embedded in a larger group $G'$ and then impose a gauge redundancy on it. For example, suppose $c$ is an $m$-cochain which we wish to take values in $C^{m}(M,\mathbb Z_q)$. It may be useful to define $c\in C^m(M,\mathbb Z)$ and then impose a gauge redundancy $c\to c+qf$, where $f\in C^m(M,\mathbb Z)$ is arbitrary. This effectively makes $c$ describe a cochain in $C^m(M,\mathbb Z / q\mathbb Z = \mathbb Z_q)$. If we further want the cochain $c$ to be closed, we then have to impose $dc=0\bmod k$. 

Finally, the dual lattice $\tilde M$ is obtained from the original $d$-dimensional lattice by a positive translation in all directions by one half of a lattice unit. A given $m$-cell on the dual (resp. original) lattice is naturally associated with the $(d-m)$-cell on the original (resp. dual) lattice it pierces. This relation is captured by the Hodge star operation $\star$ which extends to cochains: $\star : C^m(M,G) \to C^{d-m}(\tilde M,G)$, and satisfies $\star^2 \alpha = (-1)^{m(d-m)}\alpha$. 

In the Villain approach to $U(1)$ lattice gauge theory, the dynamical variables are real-valued link fields $a \in C^1(M,\RR)$ and integer-valued plaquette variables $n \in C^2(M,\ZZ)$. The link variables have the usual gauge redundancy 
\begin{equation} \label{eq:0form_redundancy} 
a \to a + d\lambda
\end{equation} 
with $\lambda \in C^0(M,\RR)$, but also shift under \emph{large} gauge transformations 
\begin{equation} \label{eq:1form_redundancy} 
a \to a + 2\pi m
\end{equation} 
with $m \in C^1(M,\ZZ)$, which causes $a$ to effectively describe a $1$-cochain in $C^1(M,\RR/ 2\pi \ZZ = U(1))$, as expected for a lattice $U(1)$ gauge field. The plaquette variable $n$ is a gauge field for these discrete shifts, and accordingly $n \to n + dm$ under such a gauge transformation. The quantity
\begin{equation}
\sum_{p\in \Sigma} n_p \in \ZZ
\end{equation}
is gauge invariant provided $\Sigma$ is a closed surface, and is interpreted as the magnetic flux through the surface $\Sigma$. Configurations where $dn \not= 0$ are similarly interpreted as \emph{monopole} configurations, since the flux through a contractible (homologically trivial) surface is equal to the sum of $dn$ on each cube enclosed by the surface. If $dn =0$ everywhere, the magnetic flux can only be non-vanishing through homologically nontrivial surfaces. In this case $n\in H^2(M,\mathbb Z)$. The continuum interpretation of $n$ in that case is that it describes the 1st Chern class of the $U(1)$ bundle.

The key ingredients for constructing our CS action is the cup product $\cup$ on the lattice and its higher generalizations (i.e. $\cup_1,\cup_2,\dots$, see below). The cup product of a $p$-cochain $\alpha$ and $q$-cochain $\beta$ is a $(p+q)$-cochain $\alpha\cup \beta$. The cup product is similar to the wedge product in de-Rham cohomology, and satisfies the Leibniz rule
\begin{equation}
d(\alpha\cup\beta) = d\alpha \cup \beta + (-1)^p \alpha \cup d\beta\,,
\end{equation} 
which one can use to establish the `summation by parts' identity
\begin{equation}
\sum d\alpha \cup \beta = (-1)^{p+1} \sum \alpha \cup d\beta\,,
\end{equation} 
where the sum is over any $(p+q+1)$-cycle. A crucial feature of the cup product is that, unlike the wedge product, it is not graded commutative. Instead, cochains (anti)commute \emph{up to} higher cup products, 
\begin{align} \label{eq:cupidentity} 
\alpha \cup \beta - (-1)^{pq}\,  \beta \cup \alpha & = (-1)^{p+q+1} \Big[ d(\alpha\cup_1\beta)\\
&- d\alpha \cup_1 \beta - (-1)^p\, \alpha \cup_1 d\beta \Big]\,. \nonumber
\end{align}
The $\cup_i$ product of a $p$-cochain $\alpha$ and $q$-cochain $\beta$ is a $(p+q-i)$-cochain. The higher cup products were introduced in \cite{Steenrod1947ProductsOC} for triangulations and have appeared in various places in the physics literature in the study of anomalies and topological phases of matter~\cite{Kapustin:2013qsa,Kapustin:2014gua,Kapustin:2014zva,Chen:2018nog,DeMarco:2019pqv,Chen:2019wlx,Tata:2020qca,Chen:2021ppt,Chen:2021xks}. The higher cup products are neither graded commutative nor associative. 

The (higher) cup products have a simple geometric interpretation. Roughly speaking ordinary cup products have to do with `generic' intersections,\footnote{What is meant by this statement is that $p$-forms (i.e. $p$-cochains) are associated by Poincar\'e duality to codimension-$p$ surfaces (see Sec.~\ref{sec:backgroundfields} for more details). So for a pair of cochains $\alpha$ and $\beta$, the cup product $\alpha\cup \beta$ measures the intersection of the lines, surfaces or hypersurfaces corresponding to their Poincar\'e duals.} e.g. two surfaces intersecting at a line in three dimensions, a line intersecting a surface at a point in three dimensions, two lines intersecting at a point in two dimensions, etc. The higher cup products detect `non-generic' intersections, e.g.  a line-like intersection of a surface and a line in three dimensions, the point-like intersection of two lines in three dimensions, the line-like intersection of two lines in two dimensions, etc. Such non-generic intersections are natural on the lattice but can always be resolved as generic intersections (or no intersections at all) in the continuum. We show some examples to illustrate this interpretation below in Fig.~\ref{fig:geometry}. 

\begin{figure}[h!] 
\centering
\includegraphics[width=0.45\textwidth]{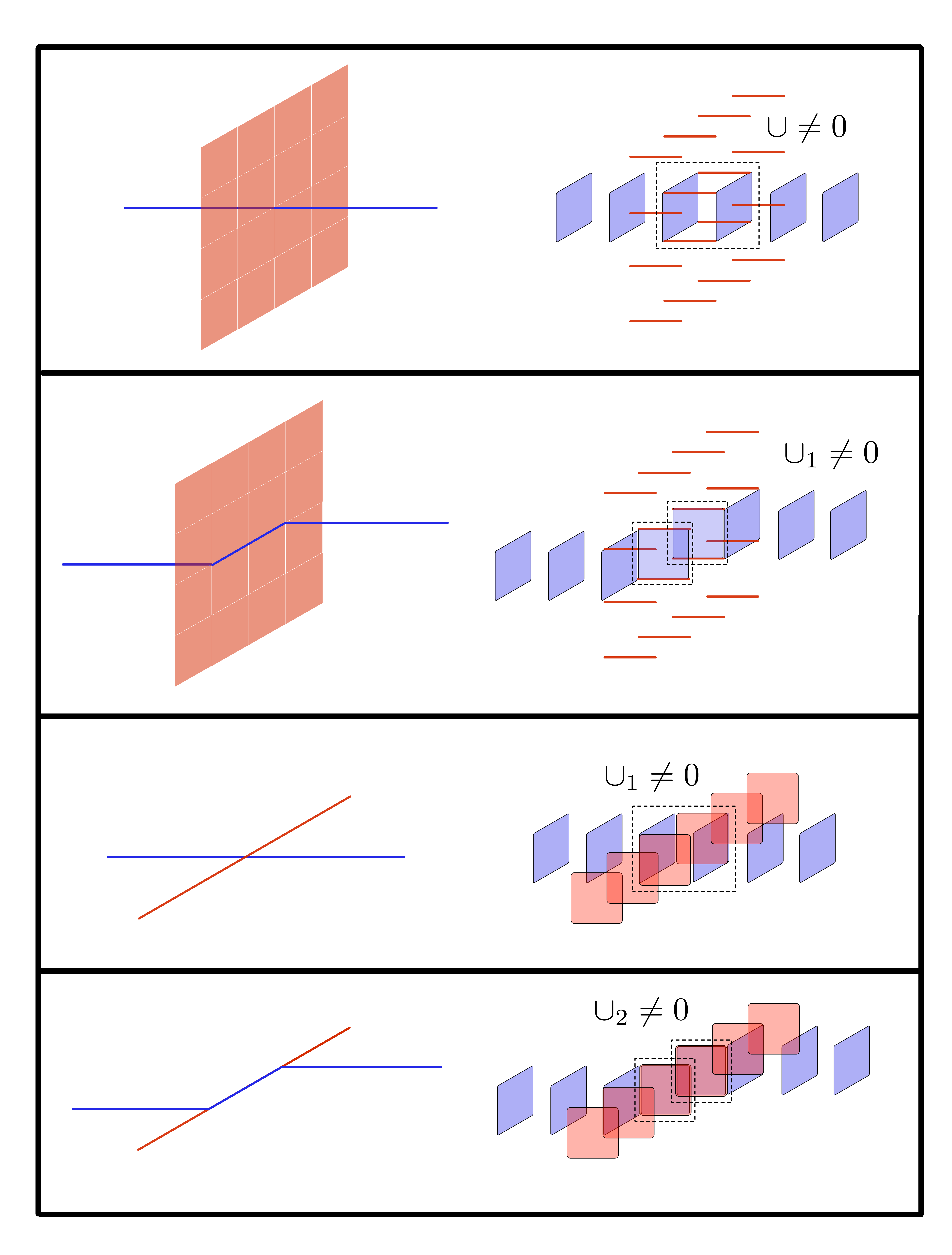}
\caption{On the left are some intersecting objects on the dual lattice, which via Poincar\'e duality correspond to the cochains on the original lattice drawn on the right. We have indicated the links or plaquettes on the original lattice which pierce or are pierced by the surfaces and lines on the dual lattice, as well as the location of some non-vanishing (higher) cup products.  } 
\label{fig:geometry}
\end{figure}

The hypercubic analog of the higher cup products have only appeared recently in the physics literature~\cite{Chen:2018nog} and were systematically defined in a  combinatorial way in~\cite{Chen:2021ppt}. Cup products on a simplicial lattice depend crucially on a choice of the branching structure, or ordering of vertices, which induces a choice of framing for each link on the lattice. The choice of branching structure is replaced by a fixed definition of the cup products in the hypercubic case---for completeness we give explicit formulas with graphical aids for (higher) cup products in App.~\ref{sec:products}.

%%%%%%%%%%%%%
\subsection{Gauge invariance and level quantization} 
%%%%%%%%%%%%%
We begin by introducing $U(1)$ gauge fields in the Villain formulation $(a,n)$ where $a\in C^1(M,\mathbb R)$ and $n\in C^2(M,\mathbb Z)$. We impose the following gauge symmetry
\begin{equation}
\begin{split}
&a\rightarrow a+d\lambda+2\pi m\\
&n\rightarrow n+dm\,,
\end{split}
\end{equation}
with $\lambda\in C^0(M,\mathbb R),\, m\in C^1(M,\mathbb Z)$.
The most naive lattice action that mimics the continuum CS term is simply
\begin{equation}
\sum_c \frac{ik}{4\pi} a\cup da,
\end{equation}
with the sum being over all cubes of the lattice, which we assume has no boundary. For now we allow the level $k$ to be arbitrary, but soon we will see that it must be quantized. The above form of the action is invariant under ordinary gauge transformations, but under large gauge transformations $a\rightarrow a+2\pi m$ (where $m$ is a $\mathbb Z$ valued 1-cochain) it shifts by
\begin{equation}
\sum_c \frac{ik}{2} \left[ m \cup da + a \cup dm + 2\pi\, m \cup dm \right]
\end{equation}
For now let us ignore the last term appearing above. After summing by parts it is clear that the first two terms can be cancelled by including additional terms in the action involving the discrete magnetic flux, 
\begin{equation} \label{eq:actionguess} 
\sum_c \frac{ik}{4\pi} \left[ a\cup da - 2\pi (a \cup n + n \cup a) \right].
\end{equation}
However, these extra terms are not invariant under ordinary gauge transformations, but shift by
\begin{equation}
-\frac{ik}{2}\sum_c d\lambda \cup n + n \cup d\lambda = \frac{ik}{2} \sum_c \lambda \cup dn + dn \cup \lambda\,.
\end{equation}
We see that the gauge variation vanishes if the discrete gauge field $n$ is flat, $dn=0$. In other words, to maintain gauge invariance we must remove dynamical monopoles from the theory. 

This can be accomplished by introducing a Lagrange multiplier field $\varphi \in C^0(M,\RR)$ and adding the term
\begin{equation} \label{eq:lagrangemultiplier} 
i\sum_c \varphi \cup dn
\end{equation} 
to the lattice action.\footnote{On a triangulation, Lagrange multiplier terms defined via a cup product fail to enforce the desired cell-by-cell constraints~\cite{Kapustin:2014gua,Thorngren:2020aph} (unless the constrained quantity is a top form), and one needs the auxiliary variables  $\varphi$ to live on the dual lattice. On the cubic lattice, there is no such requirement.} Integrating out $\varphi$ localizes the path integral on configurations for which the action \eqref{eq:actionguess} is (0-form) gauge-invariant. Equivalently, since integrating over $\varphi$ on all sites projects onto a gauge-invariant path integral weight for $a$, we should be able to write a gauge-invariant action which includes the coupling \eqref{eq:lagrangemultiplier} provided $\varphi$ itself shifts appropriately under gauge transformations. 
 
This leads us to the CS action quoted in the introduction:
\begin{equation}
\label{eq:action} 
\begin{split} 
S(a,n,\varphi) = \sum_c \frac{ik}{4\pi} \left[ a\cup da - 2\pi (a\cup n + n \cup a) \right] \\
-\frac{ik}{2}a \cup_1 dn + i \varphi \cup dn\,,
\end{split}
\end{equation}
where the term involving the $\cup_1$ product ensures that the action is invariant under gauge transformations which act by $a \to a+d\lambda$ \emph{and} $\varphi \to \varphi - k \lambda$. Under this shift, the action changes by
\begin{align}
S&(a+d\lambda ,\,n, \,\varphi -k\lambda) - S(a,n,\varphi) \\
&= -\frac{ik}{2} \sum_c  \left(d\lambda \cup n + n \cup d\lambda + d\lambda\cup_1 dn + 2\lambda\cup dn \right)\,. \nonumber
\end{align}
We now apply the cup product identity from Eq.~\eqref{eq:cupidentity} with $\alpha = n, \beta = d\lambda$, 
\begin{equation}
n\cup d\lambda + d\lambda\cup_1 dn = d\lambda\cup n - d(d\lambda\cup_1 n)\,,
\end{equation}
to find  
\begin{multline} 
S(a+d\lambda ,\,n, \,\varphi -k\lambda) - S(a,n,\varphi) \\
= -\frac{ik}{2}\sum_c \left(2d\lambda \cup n + 2\lambda\cup dn-d(d\lambda\cup_1dn) \right)  \\
= -\frac{ik}{2} \sum_c d(2\lambda \cup n-d\lambda\cup_1dn) = 0\, . 
\end{multline}
Therefore, the action in Eq.~\eqref{eq:action} is invariant under ordinary, 0-form gauge transformations. 

Now we turn to invariance under large (discrete 1-form) gauge transformations and quantization of the level $k$. Under a large gauge transformation, we have
\begin{align}  \label{eq:largegaugevariation}
S&(a+2\pi m, \, n + dm, \, \varphi) - S(a,n,\varphi)\\
&= -ik\pi\sum_c   ( m \cup n + n \cup m + m\cup_1 dn + dm \cup m ) \,. \nonumber
\end{align}
Unlike in the 0-form case, cup-product identities cannot be used to recast this as a total derivative. Moreover, the above sum can be an arbitrary integer,\footnote{The following is an example of a field configuration for which the above sum is equal to 1: take $m_{x, 3} = m_{x+\hat 3,1} = 1$ for some site $x$ and otherwise vanishing. Then the last term in Eq.~\eqref{eq:largegaugevariation} is equal to 1 on a single cube and 0 everywhere else.} so in order for the exponentiated action to be invariant we are forced to take $k$ to be an even integer, $k \in 2\ZZ$. This is the famous level quantization condition.\footnote{The level $k$ can also be an odd integer if we define the theory using an appropriate auxiliary 4d bulk (this is discussed later in Sec.~\ref{sec:thetaterm}). Although the definition of the action with auxiliary bulk will not dependent on the choice of bulk extension as long as the lattice describes a spin manifold, we are unable to construct an intrinsically 3d construction of the odd $k$ CS lattice theory. This is perhaps not surprising, because odd $k$ CS theories are spin theories, and as such they depend on the spin structure. While we believe that this construction can be used to define odd $k$ CS theories on the lattice, we mostly focus on the even-$k$ case here. }

Finally, the action is invariant mod $2\pi i$ under additional discrete shifts of the Lagrange multiplier $\varphi \to \varphi + 2\pi r$, with $r \in C^0(M,\ZZ)$. This gauge redundancy effectively makes $\varphi$ a compact scalar with radius $2\pi$. In 3d abelian gauge theory without a CS term, we could identify $\mathcal M \equiv e^{i\varphi}$ as a monopole operator, since the insertion of such an operator inserts a unit magnetic flux through Eq.~\eqref{eq:lagrangemultiplier}. However, in CS theory $\mathcal M$ is not gauge-invariant and can only exist at the endpoints of a charge-$k$ Wilson line. We return to this point in Sec.~\ref{sec:open_lines}.

To summarize, we have constructed a Chern-Simons action \eqref{eq:action} which is invariant under the following gauge redundancies on a lattice without boundary provided $k$ is an even integer: 
\begin{equation} \label{eq:gauge_transformations} 
\begin{split}
&a \to a + d\lambda + 2\pi m, \, \\&n \to n + dm, \, \\&\varphi \to \varphi -k\lambda + 2\pi r\,,
\end{split}
\end{equation}
where $\lambda \in C^0(M,\RR), m \in C^1(M,\ZZ), r \in C^0(M,\ZZ)$.

%%%%%%%%%%%%%
\section{Symmetries}
%%%%%%%%%%%%%
\label{sec:symmetries} 

We can look for 1-form symmetries by shifting $a \to a + \epsilon$ with $\epsilon \in C^1(M,\RR)$. The action shifts by
\begin{align}
\begin{split}
\Delta S = \sum_c \frac{ik}{4\pi} \Big[& \epsilon \cup da + a \cup d\epsilon + \epsilon \cup d\epsilon \\
& - 2\pi (\epsilon \cup n + n \cup \epsilon + \epsilon \cup_1 dn ) \Big] \\
= \sum_c \frac{ik}{4\pi} \Big[& d\epsilon \cup a + a \cup d\epsilon + \epsilon \cup d\epsilon \\
& - 2\pi (\epsilon \cup n + n \cup \epsilon + \epsilon \cup_1 dn) \Big] \label{eq:electricshift}
\end{split}
\end{align}
Now suppose $\epsilon = \frac{2\pi}{k} \omega$, with $\omega \in C^1(M,\ZZ)$ and $d\omega = 0\bmod k$, i.e. $\omega$ is a $\ZZ_k$ cocycle. Then we have
\begin{align}
\begin{split}
\Delta S &= -i\pi \sum_c (\omega\cup n + n \cup \omega + \omega \cup_1 dn) \\
&= -i\pi \sum_c (\omega\cup n - n \cup \omega- \omega\cup_1 dn) \text{ mod } 2\pi i \\
&= -i\pi \sum_c \left( d(\omega \cup n) - d\omega \cup_1 n \right)  = 0 \text{ mod } 2\pi i.
\end{split}
\end{align}
The first term vanishes when summed over the entire lattice, and the second term is zero mod $2\pi i$ because $d\omega=0\bmod k$ and we assume $k$ to be even. Hence, shifting the gauge field by a $\ZZ_k$ cocycle leaves the exponentiated action invariant---this is the electric 1-form symmetry of the CS theory. 

\begin{figure}[th] 
\centering
\includegraphics[width=0.3\textwidth]{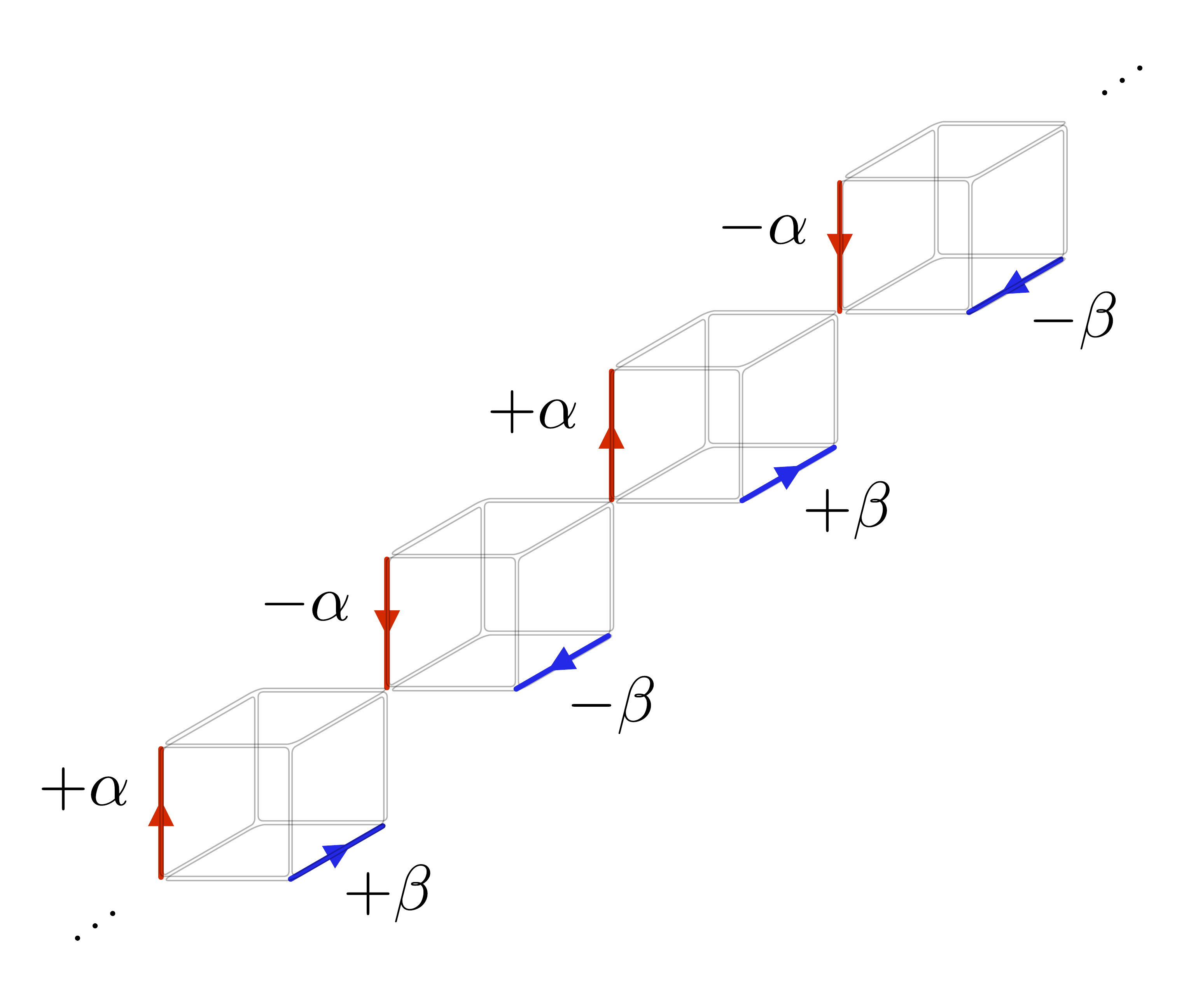}
\caption{A 1-cochain $\epsilon$ which takes values $\pm \alpha$ and $\pm\beta$ on the indicated links and 0 everywhere else satisfies $\epsilon_{f^2(\ell)} = -\epsilon_\ell$. } 
\label{fig:staggeredsymmetry}
\end{figure}

There is another interesting class of transformations that leaves the action invariant. Integrating by parts, we can rewrite the shift of the action under $a \to a + \epsilon$ as 
\begin{align}
\label{eq:staggeredshift}
\Delta S = \sum_c \frac{ik}{4\pi} \Bigg[& \epsilon \cup \left(da- 2\pi n+\frac{1}{2}d\epsilon  \right) \\
&+ \left( da - 2 \pi n+ \frac{1}{2}d\epsilon  \right) \cup \epsilon - 2\pi \epsilon \cup_1 dn \Bigg]\,. \nonumber
\end{align}
Without loss of generality we can integrate out $\varphi$ to set $dn=0$ and ignore the last term.\footnote{Alternatively we can assign $\varphi$ a compensating shift. } Then, if we can choose $\epsilon$ such that 
\begin{equation}
\sum_c \epsilon \cup X + X \cup \epsilon = 0
\end{equation}
for all 2-cochains $X$, the action is left invariant. By examining the definition of the cup product one can see that this condition is equivalent to
\begin{equation}
\sum_p X_p (\epsilon_{f^{-1}(\star p)} + \epsilon_{f(\star p)}) = 0,
\end{equation}
where $f$ is a half-unit lattice translation in the $\hat x + \hat y + \hat z$ direction and $\star p$ is the link on the dual lattice which pierces $p$. The above condition is satisfied if $\epsilon_{f^2(\ell)} = - \epsilon_\ell$ for all links $\ell$. 
%In order to satisfy the condition in Eq.~\eqref{eq:condition2} there needs to be further correlation between nearby links. An example\footnote{An inspection of the $\cup_1$ definition between a $1$-chain and a $3$-chain in Fig.~\ref{fig:cup_d} makes this choice obviously have the property that $\epsilon\cup_1 Y=0$ for any $Y$. } 
An example of such an $\epsilon$ is given in Fig.~\ref{fig:staggeredsymmetry}. Note that on a toroidal lattice the set of transformed links `wraps around' the entire lattice and consistency requires the number of lattice sites in each direction to be even. 

This extra invariance is directly related to the aforementioned zero modes which are a common feature of lattice CS constructions~\cite{Berruto:2000dp,Eliezer:1991qh,Eliezer:1992sq,Chen:2019mjw}.\footnote{The extra zero modes appear whenever $p_1+p_2+p_3=\pi\bmod 2\pi$ where $p_\mu$ are the quasi momenta of the gauge field $a$. On the other hand the change $a\rightarrow a+\epsilon$ is a symmetry as long as $\epsilon$ is odd under the diagonal translation in all directions. This means that $\epsilon$ consists precisely of modes for which $p_1+p_2+p_3=\pi\bmod 2\pi$.}  Viewed as a symmetry, it is natural to ask which operators carry charge under the staggered shifts of $a$, and which operators are neutral.
%\footnote{We use the term `symmetry' here to mean an invariance of the exponentiated action. The staggered nature of the shifts implies that there do not exist topological operators which generate the symmetry in the sense of Ref.~\cite{Gaiotto:2014kfa}.} 
One can quickly convince themselves that ordinary Wilson loops of any size transform under the staggered symmetry. This can be used to conclude that such ordinary Wilson loops have identically vanishing expectation values. To see this, we start with a Wilson loop on a single plaquette $p$ 
\begin{equation}
W(p) = \prod_{\ell\in \partial p } e^{i a_\ell }\,,
\end{equation}
and perform the transformation $a\to a + \epsilon$, where $\epsilon_{f^2(\ell)} = - \epsilon_\ell$ and $\epsilon_\ell = \alpha$ for some $\ell \in \partial p$. This field redefinition leaves the action invariant but multiplies the single-plaquette Wilson loop by $e^{i\alpha}$. As a result, the expectation value must vanish.

Note that this is what one expects from a gauge redundancy rather than a global symmetry. A line operator charged under a 1-form gauge symmetry vanishes identically for any size loop, while a gauge-invariant, contractible line operator charged under a 1-form global symmetry only vanishes in the limit where the size of the loop goes to infinity (provided the symmetry is unbroken). In this sense the staggered symmetry behaves like a gauge symmetry.\footnote{The reason for this behavior is that this staggered symmetry cannot be spontaneously broken. This is  because it can be viewed as a continuous subsystem symmetry of an effectively one-dimensional subsystem. It may be interesting to explore in more detail the relation of this symmetry structure to known subsystem symmetries.  } Note that on the one hand, adding a Maxwell term lifts the staggered symmetry. On the other hand, the Maxwell term will not be generated in our pure CS lattice theory. 

As mentioned in the introduction, it is well-known that in continuum CS theory ordinary Wilson loops are ill-defined and require point-splitting regularization~\cite{Witten:1988hf}. Such point-splitting `frames' the Wilson line, turning it into a ribbon. The staggered symmetry associated to the extra zero modes on the lattice performs the welcome function of completely projecting out all ordinary, line-like Wilson loops. However, looking at Fig.~\ref{fig:staggeredsymmetry}, it is clear that a pair of identical Wilson loops which are displaced relative to one another by one positive lattice unit in each direction will be neutral under the symmetry transformation. Such `doubled' Wilson loops are precisely the framed, ribbon-like Wilson loops on the lattice, and make up the set of physical operators. In the next section we describe how to construct and manipulate these operators by turning on background fields for the $\ZZ_k$ 1-form symmetry. 

%%%%%%%%%%%%%
\section{Background fields and framed Wilson loops}
%%%%%%%%%%%%%
\label{sec:backgroundfields}

As we argued above, ordinary Wilson loops have vanishing expectation values (and generically, correlators). On the other hand, in continuum CS theory, Wilson lines are topological and generate a $\ZZ_k$ 1-form symmetry, whose 't Hooft anomaly is encoded in the anyonic linking relations between Wilson loops~\cite{Gaiotto:2014kfa,Hsin_2019}. In other words, in CS theory a Wilson line is both the charge and the charged object of a symmetry. The fact that the charges are conserved explains their topological nature, while the fact that lines are charged objects explains why their linking is nontrivial. Our task is then to find Wilson loops which do not vanish, but are topological and correspond to charges of the 1-form electric $\ZZ_k$ symmetry. As we will see, such loops will end up being framed Wilson loops, or ribbons.

To discover these Wilson loops we will couple the theory to background gauge fields for the $\ZZ_k$ 1-form symmetry. The gauge fields of a 1-form symmetry are 2-forms (or rather 2-cochains), and since the symmetry in question is discrete the gauge field must be flat and so is really a 2-cocycle. Such an object can be described at the cochain level by an integer valued field $B$ living on plaquettes of the lattice (i.e. $B\in C^2(M,\mathbb Z))$, with a gauge symmetry 
\begin{equation}\label{eq:B_gauge}
B\rightarrow B+dV+kL\;,
\end{equation}
where $V$ is an arbitrary integer-valued field living on links, and $L$ is an arbitrary integer valued field living on plaquettes. Further we impose $dB=0\bmod k$, which implements flatness of the $B$ field.\footnote{The flatness is only meaningful modulo $k$ as the gauge field $B$ is meant to represent a $\ZZ_k$ gauge field. The failure of the $B$ field to be flat is associated with a monopole operator at the cube on which $dB\ne 0$. Such a monopole operator lies at the endpoint of a charge $k$ topological Wilson line.} Hence $B$ is really a representative of $H^2(M,\ZZ_k)$. 

To any such field $B$ on the lattice there corresponds a network of lines defined on the dual lattice. This correspondence is called Poincar\'e duality, and goes as follows. Imagine a simple contour on the dual lattice $\tilde C$. Such a contour pierces some collection of plaquettes on the original lattice. To this contour we can associate a 1-cochain $[\tilde C] \in C^1(\tilde M,\ZZ)$ on the dual lattice which counts the oriented number of times a dual link is traversed by $\tilde C$, and a 2-cochain $\star[\tilde C] \in C^2(M, \ZZ)$ on the original lattice which counts the oriented number of times a plaquette is pierced by $\tilde C$. 

\begin{figure}[h] 
\centering
\includegraphics[width=0.45\textwidth]{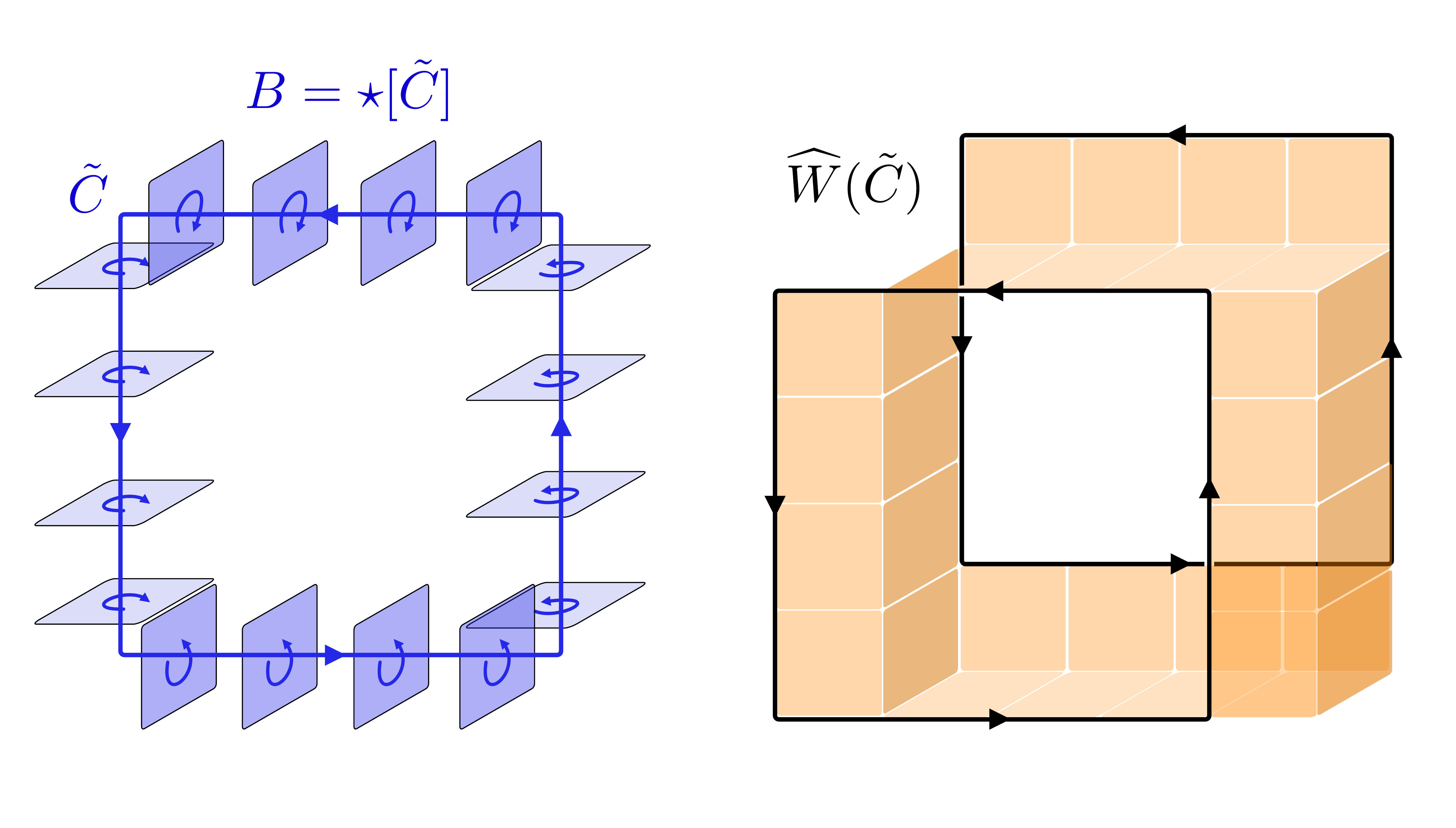}
\caption{A framed Wilson loop defined by turning on a background gauge field $B$ for the $\ZZ_k$ 1-form symmetry which is Poincar\'e dual to the curve $\tilde C$ on the dual lattice. } 
\label{fig:framed_wilson}
\end{figure}

Now take $B=\star[\tilde C]$ (see left of Fig.~\ref{fig:framed_wilson}). Such a $B$ configuration clearly has the property that it is flat $dB=0$ if the contour is closed $\partial \tilde C =0$. Alternatively we may have that $dB$ is a multiple of $k$, in which case the contours in $\tilde C$ can end in multiples of $k$. An arbitrary 2-cochain $B$ can be described by a collection of contours, which we simply denote by $\tilde C$, where $B=\star[\tilde C]$. 

Next, imagine that we employ a gauge transformation $B\rightarrow B+dV$. It takes little thought to convince oneself that $dV$ corresponds to inserting arbitrary contours on the dual lattice which are contractable, i.e. contours $\tilde\Gamma$ for which $\tilde \Gamma = \partial\tilde\Sigma$. This means that we can use the gauge freedom to deform the set of contours corresponding to $B$ as we wish. Finally, the gauge freedom that $B\rightarrow B+kL$ simply tells us that inserting non-contractable or open contours does not change anything as long as they come in multiples of $k$. This is just a statement that $k$ contours can annihilate. In fact all of these properties are exactly features of a collection of lines which measure the 1-form $\ZZ_k$ charge.\footnote{The statement that a $\ZZ_k$ symmetry is free from 't Hooft anomalies (i.e. the theory is completely background gauge-invariant) means that only the $\ZZ_k$ homology of the lines are important. In CS theory the anomaly implies that the corresponding lines (or rather strips, as we shall see) are only topological up to linking, intersections, and topological twists. }

%(such that it does not identically vanish) by coupling the theory to a background field for the 1-form $\ZZ_k$ symmetry. A generic background field is equivalent to inserting a network of symmetry defects, which are just the topological Wilson loops. 

In a conventional situation, turning on a background field $B = \star[\tilde C]$ would be equivalent to inserting topological defects, or symmetry generators, supported on the lines $\tilde C$ on the dual lattice. In the present case, we expect such operators to be topological Wilson lines, which live on the original, rather than dual, lattice. To see how this works out, we couple Eq.~\eqref{eq:action} to a background gauge field $B$ for the $\ZZ_k$ 1-form symmetry. The $1$-form global symmetry transformation $a\rightarrow a+\frac{2\pi}{k}\omega$ can be promoted to a background gauge redundancy via the minimal substitution $n \to n + \frac{1}{k}B$. This is quite natural because physically, coupling the theory to a background field for the 1-form symmetry relaxes the quantization condition \eqref{eq:quantized_flux} to allow for fractional fluxes. The coupling to background fields is
\begin{multline} \label{eq:1-form_bgd} 
S(a,n,\varphi ;B) = \sum_c \frac{ik}{4\pi}\Big[ a\cup da - 2\pi a \cup \left(n +\frac{1}{k}B\right)  \\
 - 2\pi \left(n + \frac{1}{k}B\right)\cup a - 2\pi a \cup_1 \left(dn+ \frac{1}{k}dB \right) \Big]  \\
 +i \varphi\cup\left( dn + \frac{1}{k}dB \right) + i\pi B \cup_1 n\,. 
\end{multline}
The dynamical fields $a$ and $n$ shift under these background gauge transformations as
\begin{equation}
a \to a + \frac{2\pi}{k}V, \ n \to n - L\,.
\end{equation}
Note that the last term in Eq.~\eqref{eq:1-form_bgd} does not arise from any minimal coupling, but plays an important role. We will come back to it in a moment. 

First, let us check that the action remains invariant under dynamical gauge transformations even in the presence of background fields. Repeating the analysis around Eq.~\eqref{eq:largegaugevariation}, under dynamical gauge transformations the action coupled to $B$ has an additional shift by
\begin{multline}\label{eq:gaugevariation2}
 -i\pi \sum_c ( B \cup m + m\cup B + m\cup_1 dB - B\cup_1 dm )  \\
 = -i \pi \sum_c (d(B \cup_1 m) - dB\cup_1 m + m\cup_1 dB)  \\
 = 0 \text{ mod } 2\pi i, 
\end{multline}
where we used the cup product identity Eq.~\eqref{eq:cupidentity} with $\alpha = B$ and $\beta = m$ and the last equality follows from the fact that $dB = 0 \text{ mod } k$. So the exponentiated action coupled to background fields is gauge-invariant. This means that turning on a particular background has the effect of inserting some collection of gauge-invariant operators.

Now, we focus on a configuration $B=\star[\tilde C]$ for some single closed contour $\tilde C$ on the dual lattice, for example the one on the left side of Fig.~\ref{fig:framed_wilson}. Plugging this into the action in Eq.~\eqref{eq:1-form_bgd} (ignoring the last term for the moment), we see that such an insertion involves the terms $e^{\frac{i}{2}\sum_c (a\cup B+B\cup a)}$. A quick reference to Fig.~\ref{fig:cup_b} in the appendix reveals that this corresponds to the insertion of two charge $1/2$ Wilson lines on the original lattice, offset by a diagonal shift. These are represented by the black lines in the right of Fig.~\ref{fig:framed_wilson}. But such Wilson lines have improperly quantized coefficients and hence are not invariant under large gauge transformations. To make them gauge invariant, we need to connect them with a surface built out of the discrete variable $n$ on plaquettes lying between the two fractionally charged Wilson lines. This is exactly what the final term $e^{-i\pi \sum_c B \cup_1 n}$ in Eq.~\eqref{eq:1-form_bgd} accomplishes. We can identify the resulting operator as a framed Wilson line, which is topological by virtue of background gauge invariance. Due to the framing, these Wilson lines are really `strips' or `ribbons,' but are defined via a single curve $\tilde C$ on the dual lattice, 
\begin{align}
\widehat W(\tilde C)\equiv e^{\frac{i}{2}\sum\limits_c a \cup \star[\tilde C]} e^{\frac{i}{2}\sum\limits_c \star [\tilde C] \cup a } e^{-i\pi \sum\limits_c \star[\tilde C] \cup_1 n }\,.
\end{align}

\subsection{'t Hooft anomaly for $\ZZ_k$ 1-form symmetry}
%%%%%%%%%%%%%

Now we turn to background gauge transformations. As discussed above, these gauge transformations have the effect of adding contractible loops, or lines in multiples of $k$, to the network of symmetry defects. Invariance under such transformations implies that the corresponding symmetry operators are completely captured by their $\ZZ_k$ homology. Failure to maintain full gauge invariance indicates an 't Hooft anomaly, and a more detailed dependence of correlation functions on the topology of the symmetry defect network~\cite{Gaiotto:2014kfa}. Under a background gauge transformation the action shifts by
\begin{multline}
S(a +\frac{2\pi}{k}  V,\, n \to n-L,\, \varphi; B + dV + k L)- S(a,n,B)\\ 
= \sum_c \frac{ik}{4\pi} \Bigg[  \frac{2\pi}{k}V \cup da - \frac{2\pi}{k}dV\cup a - \left(\frac{2\pi}{k}\right)^2 dV\cup V  \\
  - \frac{2\pi}{k}V\cup \left(2\pi\, n+\frac{2\pi}{k}B  \right) -  \left(2\pi \, n+\frac{2\pi}{k}B \right)\cup \frac{2\pi}{k}V     \\
- \frac{2\pi}{k}V\cup_1 \left(2\pi \, dn + \frac{2\pi}{k}dB \right) \Bigg]  \\
+ i \pi \, dV\cup_1 (n-L) - i\pi \, B \cup_1 L \,. 
\end{multline}
Dropping total derivatives and multiples of $2\pi i$, the variation simplifies to 
\begin{align}
\begin{split}
\sum_c &-\frac{2\pi i}{2k}\left( dV\cup V +V\cup B + B\cup V \right)\\
&- i \pi \left( (B+dV)\cup_1 L + V \cup_1 \frac{1}{k}dB \right) \\
&- i \pi \left( V\cup n + n \cup V + V\cup_1 dn - dV \cup_1 n \right)   \,.
\end{split}
\end{align}
Note that the first two lines only involve background fields---they encode the anomaly of the $\ZZ_k$ symmetry, or obstruction to gauging. The last line can be rewritten, mod $2\pi i$, as 
\begin{align}
\begin{split}
- i \pi (V\cup n - n \cup V + dV \cup_1 n - V\cup_1 dn )
 \\
 = -i\pi\, d(V\cup_1 n)\,,
\end{split}
\end{align}
which is a total derivative. Hence, all terms involving dynamical fields drop out and we are left with the anomaly,
\begin{multline} \label{eq:anomaly}
S_{\text{anomaly}}(B,V,L) \\
= \sum_c -\frac{2\pi i}{2k}\left( dV\cup V +V\cup B + B\cup V \right)\\
- i \pi \left( (B+dV)\cup_1 L + V \cup_1 \frac{1}{k}dB \right)\,.
\end{multline}
Note that despite our working with a $\ZZ_k$ symmetry the anomaly displays $\ZZ_{2k}$-valued terms, as well as $\ZZ_2$-valued terms (recall $dB \in k\ZZ$) which are absent in the standard continuum analysis (see App.~\ref{sec:pontryagin}). In fact we will see that this $\ZZ_{2k}$ structure leads to the correct topological spin of framed Wilson loops. As is usually the case with anomalies, one can cancel some of the above terms by using local counter-terms involving background fields. In the present case we are limited to terms involving higher cup products such as $B \cup_1 B$ and $B\cup_2 dB$. The fact that a genuine anomaly remains is made clear by providing a four-dimensional anomaly inflow action \eqref{eq:SPT}, which we discuss later in Sec.~\ref{sec:thetaterm}. 

\begin{figure}[h] 
\centering
\includegraphics[width=0.35\textwidth]{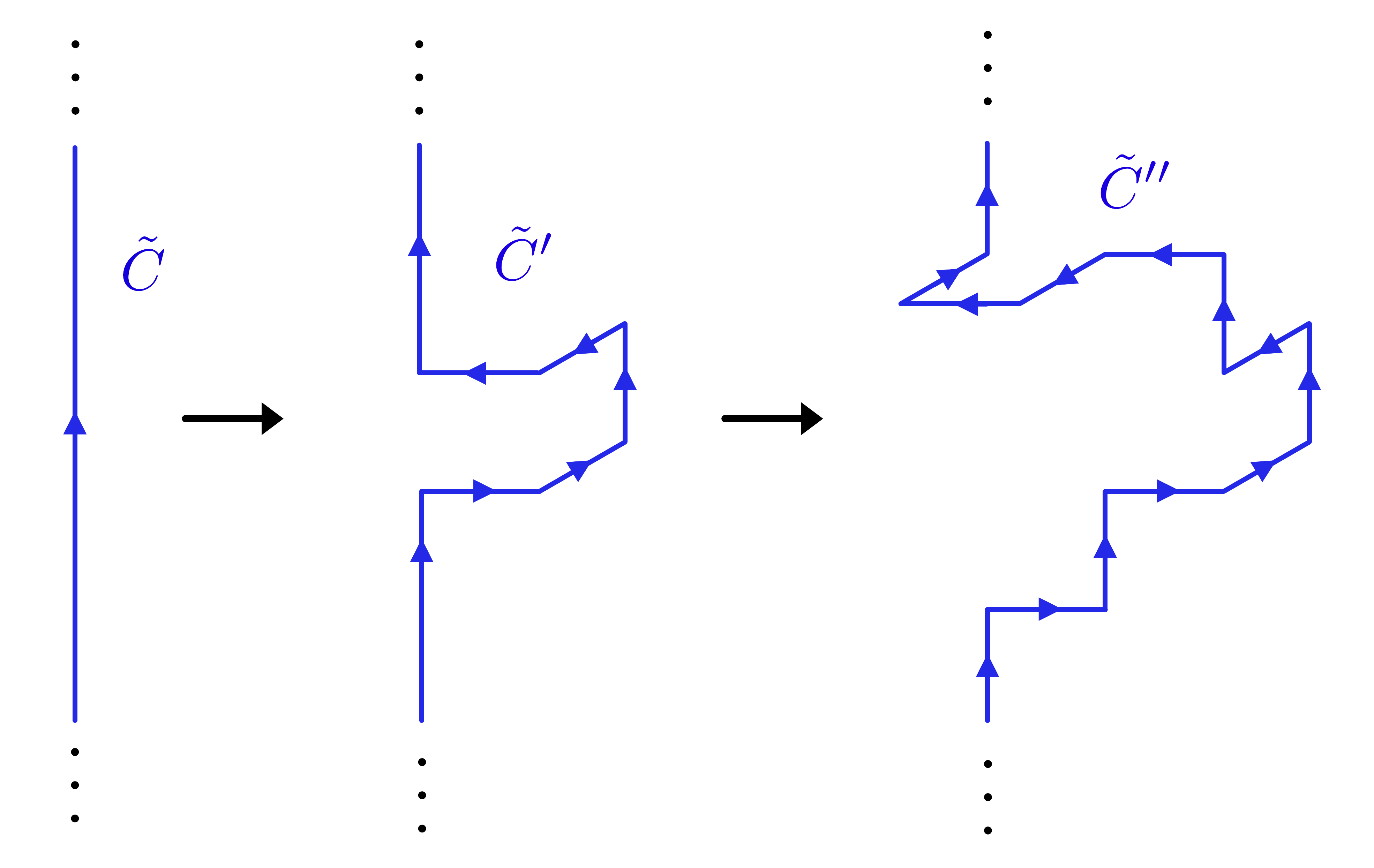}
\caption{Background gauge transformations can be used to deform a Wilson line.}\label{fig:deformation}
\end{figure}

Background gauge transformations can be used to compare correlation functions of  Wilson loops as they are topologically deformed.\footnote{To be very explicit, due to the invariance of the measure over the dynamical fields under redefinitions $a \to a + \frac{2\pi}{k}V, n \to n-L$, we have
\begin{align}
\begin{split}
\int \mathcal Da\mathcal Dn\, & e^{-S(B+dV+kL,a,n)} \\
&= \int \mathcal Da\mathcal Dn \, e^{-S(B+dV+kL,a+\frac{2\pi}{k}V,n-L)}  \\
&=e^{-S_{\text{anomaly}}(B,V,L)} \int \mathcal Da\mathcal Dn\, e^{-S(B,a,n)} \,. \end{split}
\end{align}
} To illustrate this, let us start with a straight Wilson loop and perform a set of gauge transformations $B\rightarrow B+dV$ to deform its shape. We first start with $V$ such that $V\cup B+B\cup V=0$ for all cubes $c$ such that the anomaly Eq.~\eqref{eq:anomaly} vanishes. In Fig.~\ref{fig:deformation} we show examples of such transformations, which indicate the topological nature of our framed Wilson loops.

\begin{figure}[h] 
\centering
\subfigure[]{\includegraphics[width=0.4\textwidth]{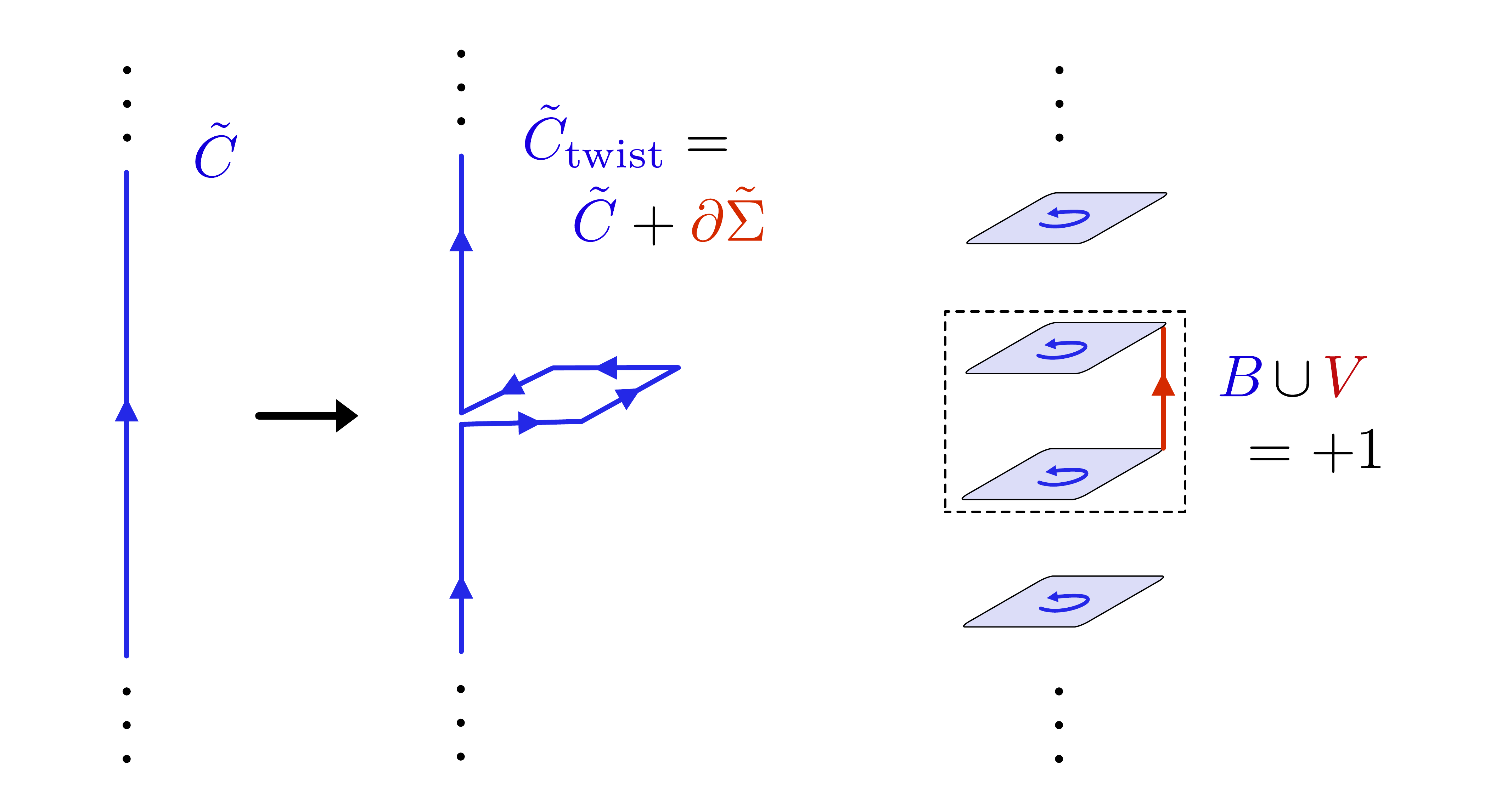} }
\subfigure[]{\includegraphics[width=0.4\textwidth]{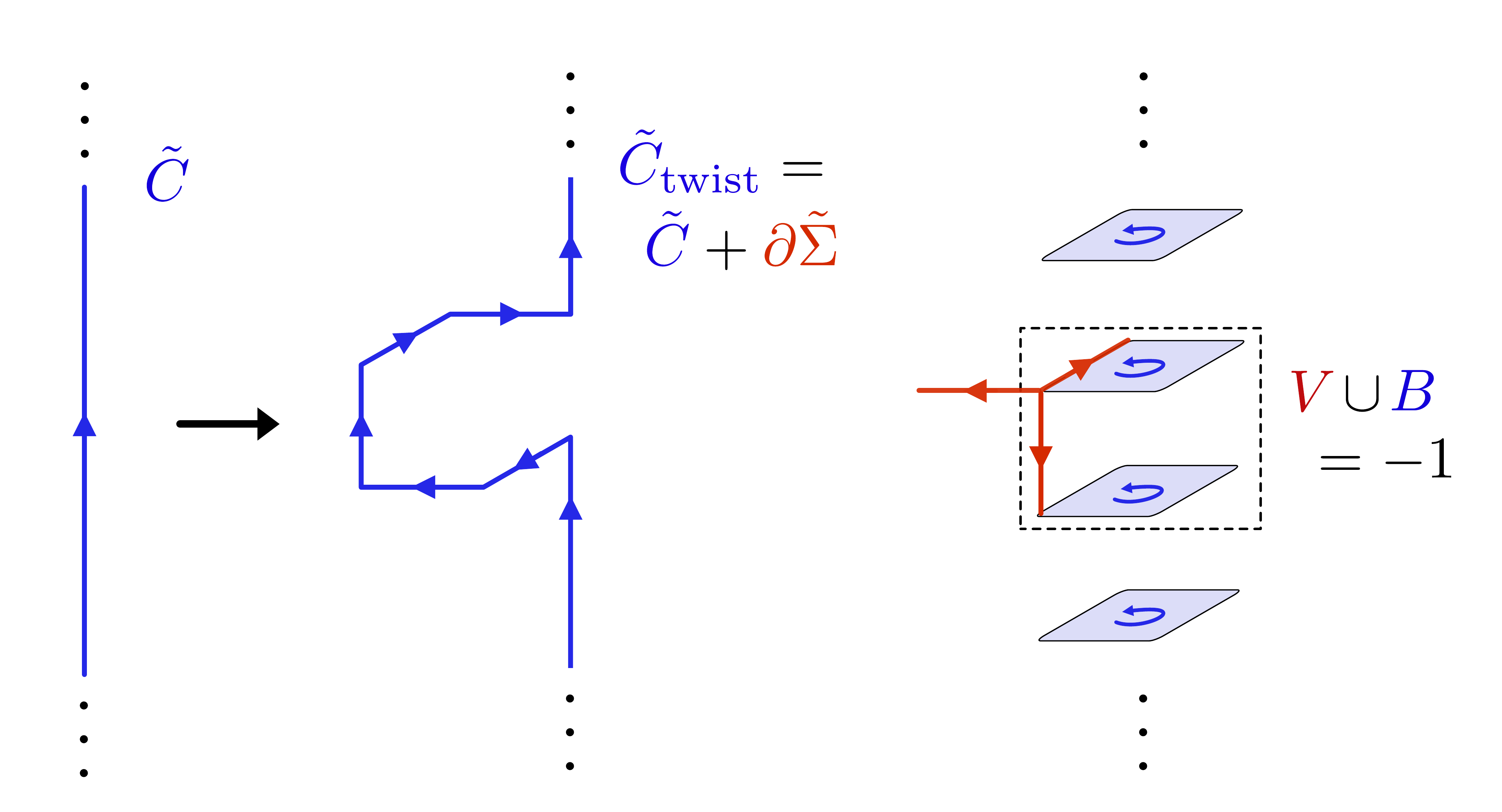} }
\caption{Twisting a straight Wilson loop $\tilde C \to \tilde C_{\text{twist}}$ introduces a $\ZZ_{2k}$ phase. The straight and twisted configurations are related via background gauge transformations, $B \to B+ dV$ where $B = \star[\tilde C]$ and $V = \star[\tilde\Sigma]$. In (a) the anomaly picks up a contribution from a single cube for which $B\cup V= +1$. Note that the anomalous phase is not coming from the self-intersection of the curve $\tilde C_{\text{twist}}$ (for this reason, we resolved the intersection point). In (b) the anomalous phase comes from a single cube for which $V\cup B = -1$. }
%\caption{Another example of adding a twist to a straight Wilson loop. In this case the anomaly picks up a contribution from a single cube for which $V\cup B = -1$. }
\label{fig:twist}
\end{figure}

Now we consider transformations which deform the Wilson loop in such a way that the anomaly induces a $\ZZ_{2k}$ phase. Examples of such transformations are depicted in Fig.~\ref{fig:twist} (see also Fig.~\ref{fig:topologicalspin}). Let us call this new loop $\tilde C_{\text{twist}}$. It follows from the anomaly that 
\begin{equation}
\langle \widehat W(\tilde C_{\text{twist}}) \rangle = e^{\pm \frac{2\pi i}{2k}} \ \langle \widehat W(\tilde C) \rangle \, ,
\end{equation}
which indicate that the contours $\tilde C_{\text{twist}}$ are indeed twisted, or in other words have non-trivial self-linking with respect to our framing. We can further identify this minimal phase resulting from twisting as the fractional $\frac{1}{2k}$ spin of an anyon~\cite{Witten:1988hf,Kitaev:2006lla}.

\begin{figure}[h] 
\centering
\includegraphics[width=0.48\textwidth]{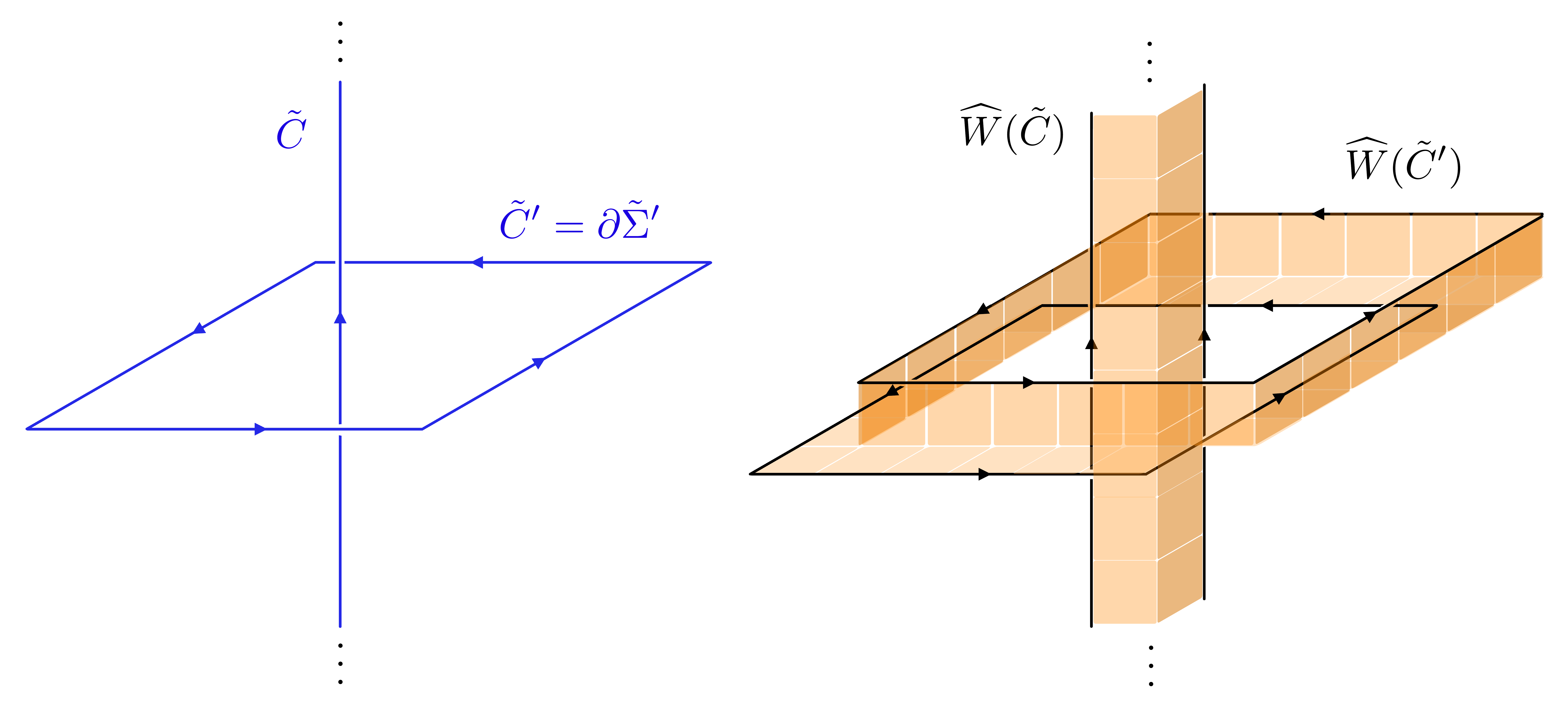}
\caption{Large framed Wilson loops with non-trivial linking. In the absence of additional twists, large non-intersecting loops yield linking-dependent $\ZZ_k$ phases. }
\label{fig:linking}
\end{figure}

Finally let us now insert a contractable $\tilde C'$ Wilson loop linking the original one $\tilde C$. This corresponds to a gauge transformation $V$ which is unity on all links pierced by the surface $\tilde \Sigma'$ whose boundary is $\tilde C'$ (see Fig.~\ref{fig:linking}). A little thought reveals that \emph{both} $V \cup B$ and $B\cup V$ are $+1$ for a single cube. Hence the anomaly induces a phase $e^{-2\pi i/k}$, so that
\begin{equation}
\langle \widehat W(\tilde C)\widehat W(\tilde C') \rangle =e^{\frac{2\pi i}{k}}\langle \widehat W(\tilde C)\rangle \langle \widehat W(\tilde C')\rangle\,.
\end{equation}
This reproduces the familiar linking relation one expects from the continuum---indeed, correlation functions of loops which are sufficiently large and far apart will yield linking-dependent $\ZZ_k$ phases. 

\subsection{Open Wilson lines and monopole operators}
%%%%%%%%%%%%%
\label{sec:open_lines} 

\begin{figure}[th] 
\centering
\includegraphics[width=0.3\textwidth]{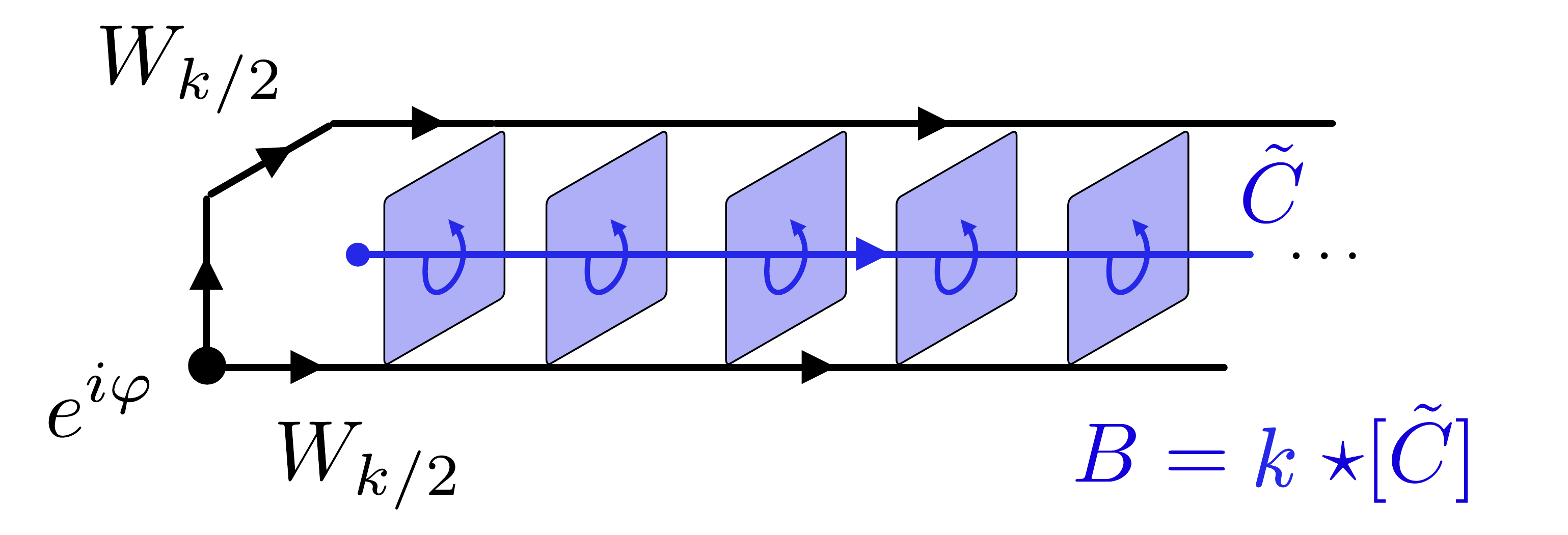}
\caption{A monopole operator attached to the endpoint of a framed Wilson line with charge $k$. } 
\label{fig:open_line}
\end{figure}

Coupling to background fields for the 1-form symmetry also gives us a way of constructing gauge-invariant monopole operators, which must be attached to Wilson lines of the appropriate charge. In particular, we can take a background field configuration which is pure gauge, $B = k \star[ \tilde C]$, which roughly corresponds to a charge-$k$ Wilson line with boundary, $\partial \tilde C \not = 0$. This activates all terms in Eq.~\eqref{eq:1-form_bgd} except for the final one. The resulting operator is shown in Fig.~\ref{fig:open_line}, and consists of two charge $k/2$ Wilson lines emanating from a single monopole operator. There is no magnetic ribbon connecting the two Wilson lines because they each have integer charge. 

Although this operator appears to be non-trivial, the fact that it corresponds to a pure-background-gauge configuration implies that it at most has contact interactions encoded in the last term of the anomaly \eqref{eq:anomaly}. Relatedly, the 1-form charge of the open Wilson line is trivial at long distances (i.e. ignoring intersections), and a straight open Wilson line such as the one in Fig.~\ref{fig:open_line} can be topologically contracted to a point. Correspondingly, there are no genuinely non-trivial monopole operators in CS theory, nor is there a faithfully-acting $U(1)$ magnetic symmetry as in pure 3d Maxwell theory in the absence of dynamical monopoles. 

That there is no $U(1)$ magnetic symmetry is also made clear by the fact that the would-be symmetry generators
\begin{equation}
\prod_{p\in\Sigma} e^{i \alpha n_p}
\end{equation} 
are trivial operators (here $\alpha$ is an arbitrary angle). Such an operator can be completely removed by an appropriate field redefinition of $a$ and $\varphi$ (or just $\varphi$ in the case that $\Sigma$ is a boundary). 

%%%%%%%%%%%%%
\subsection{Comments on zero and near-zero modes}\label{sec:comments_on_zm}
%%%%%%%%%%%%%
\label{sec:zero_modes} 

We close this section with some comments on the presence of the zero and near-zero modes which are generic for Chern-Simons discretizations and which have been studied in many works cited in the Introduction. These zero modes arise as a result of an exact symmetry of the action (see Fig.~\ref{fig:staggeredsymmetry} and the discussion around it) which we refer to as staggered symmetry. Though we have seen that the exact zero modes simply project out certain operators, one may worry that \emph{near}-zero modes could affect correlators of the surviving operators and betray the existence of the gapless sector.\footnote{We thank Max Metlitski for raising this question.} 

However, as we already explained, the staggered symmetry completely eliminates all operators which are charged under it. This includes the naive Wilson loops and a more general class of operators such as 
\begin{equation}
e^{i\alpha \sum_{p\in S}[(da)_p+2\pi n_p] }, \quad  \alpha\in \mathbb R\,,
\end{equation}
where the sum is over plaquettes belonging to some open surface $S$ on the lattice (as discussed in the previous section, summing over a closed surface would yield a trivial operator). Moreover, the Wilson lines which survive the staggered symmetry are completely topological with correlation functions dictated by the 1-form symmetry and its 't Hooft anomaly. So even if the near-zero mode sector is physical, it is completely decoupled from the Wilson strips. 

One may wonder whether there exist, aside from the topological Wilson lines, any operators which do not vanish due to the staggered symmetry, are non-trivial, and could activate these near-zero modes. The answer is no---the only other class of gauge- and stagger-invariant operators can be written as\footnote{This operator can be thought of as the generator of staggered-invariant field strength correlation functions.}
\begin{equation}\label{eq:H_term}
e^{i \sum\limits_c H \cup (da-2\pi n) + (da-2\pi n) \cup H}
\end{equation}
for some real 1-cochain $H$. However, up to a local counter-term (see below), this can be completely removed by shifting $a \to a + \frac{4\pi}{k} H$. We can therefore conclude that apart from projecting out unframed operators, the zero and near-zero modes do not affect any correlation functions. 

We make a brief comment that the counter-term we mention above is not completely removable and contains information on universal contact terms in the continuum CS theory. Namely the operator \eqref{eq:H_term} has a continuum analog as $\exp(\frac{i}{2\pi}\int W\wedge da)$, where now $W$ plays the role of $H$, up to normalization. This operator generates all correlators of the field-strength $da$, which in CS theory are pure contact terms. However, because of the flux quantization of $da$, $W$ can be viewed as a $U(1)$ gauge field. This constrains the possible counter-terms which are allowed, rendering some of the contact terms ``physical''~\cite{Closset:2012vp}.\footnote{The meaning of the word ``physical'' is as follows. In the continuum, contact terms are typically deemed unphysical because they are ambiguous. To explain this, let us pick our favorite regularization of the QFT, and consider the generating functional containing local classical sources for all operators, which we collectively label as $J$. As we flow to an intermediate energy scale where our QFT lives, we generate infinitely many local terms consistent with all the symmetries involving $J$ only. These local terms will induce contact contributions to the correlation functions. The precise coefficients of these contact terms are ambiguous, as they depend on the details of the UV completion. In the IR, this is reflected in the ability to adjust local counter-terms. It is for this reason that one says contact terms are not ``physical'' or are ``ambiguous.'' However note that contact terms of a given regulated theory, such as a lattice theory, are not ambiguous at all. Nevertheless they generically, up to possible subtleties discussed in Ref.~\cite{Closset:2012vp}, have no meaning in the IR theory.}

%%%%%%%%%%%%%
%{\bf Anomaly inflow and a 4d $\theta$ term.} 
%%%%%%%%%%%%%
\section{4d theta term and anomaly inflow} 
\label{sec:thetaterm} 

In this section we show that our CS action \eqref{eq:action} can be obtained from a particular definition of the 4d theta term on a lattice with boundary and $\theta =2 \pi k$. There are two perspectives on defining CS theory via some auxiliary bulk. One is that we define the value of the 3d Chern-Simons action by extending each field configuration into a bulk and computing the action there. Different extensions of a given 3d field configuration must yield the same action. Such an extension exists for every field configuration, but a fixed choice of bulk manifold may admit an extension of one class of field configurations but not another (for example, if they differ by global fluxes).  

In the current context we work with a fixed 4d lattice with boundary, which has the topology of $T^2 \times D$. In other words we define a bulk theory on a fixed manifold such that it reproduces Chern-Simons theory on the boundary with no bulk-dependence. We will see that this is only possible if $k$ is even. The theta term we consider is
\begin{align} \label{eq:theta} 
S_\theta(a,n,b) &= \sum_h \frac{i\theta}{8\pi^2} (da -2\pi n)\cup (da-2\pi n) \\
& -\frac{i\theta}{4\pi}(da-2\pi n)\cup_1 dn +i \left(\frac{\theta}{2\pi} a+b\right)\cup dn \,, \nonumber
\end{align}
where $b \in C^1(X,\RR)$ is a Lagrange multiplier imposing the no-monopole constraint and the sum is over all hypercubes $h$ of the 4d lattice $X$. The $\cup_1$ product between $2$- and $3$-cochains is defined in App.~\ref{sec:products}. 

This definition of the theta term differs from the one presented in Refs.~\cite{Sulejmanpasic:2019ytl,Anosova:2022cjm} in two ways. First, the Lagrange multiplier $b$, which should be interpreted as the magnetic gauge field, lives on the original lattice and not the dual lattice. Second, the action includes additional terms involving $dn$, which vanish upon integrating out $b$. These two modifications lead to certain desirable features---in particular, the above action \emph{density} is 0- and 1-form gauge invariant provided
\begin{align} \label{eq:witten_effect} 
b \to b - \frac{\theta}{2\pi} (d\lambda+2\pi m)\,.
\end{align}
This means we can easily study the theory on a manifold with boundary. In addition, the gauge field $b$ also has its own magnetic $U(1)$ gauge symmetry 
\begin{equation}
b \to b + d\beta + 2\pi s
\end{equation}
with $\beta \in C^0(X,\RR)$ and $s\in C^1(X,\ZZ)$. The magnetic gauge field transforms under electric gauge transformations due to the Witten effect~\cite{Witten:1979ey}. Owing to the fact that $b$ lives on the original lattice and not the dual lattice, these electric gauge transformations are perfectly local and do not require `splitting' the charge between neighboring links as in~\cite{Sulejmanpasic:2019ytl,Anosova:2022cjm}.\footnote{The attractive features of this theta term in the `electric' variables come at the cost of making the dual `magnetic' description (obtained by applying Poisson resummation to $n$) more involved, but still possible to perform. The dual theory will likely be non-ultralocal and to restore exact electric-magnetic duality one needs to appropriately modify the theory similarly to what was done in~\cite{Anosova:2022cjm}.}

Rewriting the action using the cup product identity Eq.~\eqref{eq:cupidentity}, we find that most terms are total derivatives:
\begin{multline} \label{eq:theta_2} 
S_{\theta}(a,n,b) =  \\
\sum_h  \frac{i\theta }{8\pi^2} d\Big[ a \cup da - 2\pi a \cup n - 2\pi n \cup a - 2\pi a\cup_1 dn \Big]  \\
 + \frac{i\theta}{2} (n\cup n  + n\cup_1 dn)  + i b \cup dn  \,.
\end{multline}
Now we set $\theta = 2\pi k$ with $k \in \ZZ$. On a lattice without boundary (where $X \simeq T^4$), this reduces to 
\begin{equation} 
S_{\theta = 2\pi k}(a,n,b) = \sum_h i k \pi (n\cup n + n \cup_1 dn) + i b \cup dn \,.
\end{equation}
If we integrate out $b$ to explicitly enforce the no-monopole constraint, the second term vanishes and on a periodic lattice $\sum_h n\cup n$ evaluates to an \emph{even} integer~\cite{Sulejmanpasic:2019ytl}. Hence the partition function of the theory with $\theta \in 2\pi \ZZ$ is equal to unity on a closed periodic lattice.\footnote{One might try to use this fact to define $U(1)_k$ CS theory with odd $k$ on the lattice through Eq.~\eqref{eq:theta_2}. However, when $k$ is odd the bulk partition function on a closed (spin) manifold is only trivial in the absence of monopoles. Here we are working with a fixed bulk lattice $X \simeq T^2 \times D$, and there exist 3d configurations which cannot be extended to $X$ without monopoles in the bulk. We however expect that there exists a bulk lattice for which the odd $k$ theory can be defined in this way. }

To make the connection to our 3d CS term \eqref{eq:action}, we now take $\theta = 2\pi k$ with $k \in 2\ZZ$ and consider the theory \eqref{eq:theta} on a lattice $X$ with boundary $\partial X$. Referring to Eq.~\eqref{eq:theta_2}, the only nontrivial term which fails to localize to the boundary is the Lagrange multiplier, unless the magnetic gauge field is restricted to be flat, $db=0$. Suppose we go further and restrict $b = d\varphi \text{ mod } 2\pi$ for some $\varphi \in C^0(X,\RR)$. This relation is gauge invariant provided $\varphi \to \varphi +\beta -k \lambda$, and we observe that with such a restriction Eq.~\eqref{eq:theta_2} reduces exactly to our CS action Eq.~\eqref{eq:action}. In other words, when $k$ is even 
\begin{equation}
S_{\theta = 2\pi k}(a,n,b = d\varphi)\Big|_X = S_{\text{CS},k}(a,n,\varphi)\Big|_{\partial X}
\end{equation}
mod $2\pi i$. 

The fact that we had to restrict the magnetic gauge field to be exact in order for the theta term to localize to the boundary has a simple interpretation in terms of Higgsing the magnetic gauge field. Indeed, we can couple $b$ to a Higgs field $\varphi$ in the Villain representation, 
\begin{equation}
\sum_\ell \frac{\kappa}{2}( (d\varphi)_\ell - b_\ell  - 2\pi u_\ell)^2
\end{equation}
where $u \in C^1(X,\ZZ)$ and $\varphi \to \varphi + \beta - k\lambda$, $u \to u -s + km$ under combined electric and magnetic gauge transformations (i.e., $\varphi$ is a dyonic Stueckelburg field). Furthermore $\varphi \to \varphi + 2\pi r$, $u \to u + dr$, as befits a compact scalar. Taking the deep Higgs limit by sending $\kappa \to \infty$ restricts $b = d\varphi$ mod $2\pi$. 

Physically, this Higgsing can be thought of as summing over all monopole worldlines in the bulk (which are really dyons due to the Witten effect). This is necessary in order to reproduce the full CS theory on the boundary for the following reason. In our 4d setup the magnetic flux variable $n$ is dual to a surface $\tilde\Sigma$ in the bulk which can end on a curve on the 3d boundary. Consider a 3d configuration where $n$ is dual to a non-contractible curve $\tilde C$, corresponding to non-vanishing flux through a 2-cycle. With our fixed bulk lattice $X \simeq T^2\times D^2$, some configurations of this type require the surface $\Sigma$ ending on $\tilde C$ to also end on a dyon worldline in the bulk. As a result, to capture all configurations on the boundary one has to sum over all dyon worldlines in the bulk with a flat weight, i.e. condense them. The condensation of dyons is known as `oblique confinement'~\cite{tHooft:1981bkw,CARDY19821,CARDY198217}.

Let us return again to the periodic 4d lattice without boundary and $b = d\varphi$,
\begin{equation} \label{eq:spt_nobgd}
S_{\theta = 2\pi k}(a,n,b=d\varphi) = \sum_h ik\pi (n\cup n + n \cup_1 dn),
\end{equation}
where we dropped the total derivative. Clearly when $k \in 2\ZZ$ the partition function is unity and this appears to be a trivial theory. In fact, it is a symmetry-protected topological (SPT) phase protected by the $U(1)$ electric 1-form symmetry of Eq.~\eqref{eq:theta} which acts by shifting $a$ by an arbitrary flat 1-form. Though seemingly trivial, the action \eqref{eq:spt_nobgd} encodes the response to background fields for this symmetry. Let us consider the $\ZZ_k$ subgroup of the electric 1-form symmetry. The SPT action coupled to a background $\ZZ_k$ gauge field $B$ reads
\begin{multline} 
S_{\text{SPT}}(B) = \sum_h \,   ik\pi \Bigg[ \left(n+\frac{1}{k}B\right) \cup \left(n+\frac{1}{k}B\right)\\
 + \left(n+\frac{1}{k}B\right) \cup_1 d\left(n+\frac{1}{k}B\right)\Bigg]\,.  \\
= \frac{2\pi i}{2k} \sum_h  \mathcal P(k\, n + B)\,,
\end{multline}
where we have introduced the Pontryagin square operation which when $k$ is even `squares' a $\ZZ_k$ cocycle to form a $\ZZ_{2k}$ cocycle ~\cite{Pontrjagin:1942,Whitehead:1949,Kapustin:2013qsa}. Explicitly, 
\begin{equation}
\mathcal P(\alpha) \equiv \alpha \cup \alpha + \alpha \cup_1 d\alpha\,,
\end{equation}
see  App.~\ref{sec:pontryagin} for some motivation behind this formula. In the present context, the combination $k\, n + B$ is a $\ZZ_k$ cocycle, and the above SPT action density takes values in $\ZZ_{2k}$. The fact that the Pontryagin square is a well-defined product in cohomology ensures that the SPT action is invariant under both dynamical gauge transformations (under which $k\, n+B \to k\, n+B +k\,  dm$) as well as background gauge transformations (under which $k\, n + B \to k\, n + B + dV$). 

We can further simplify the SPT action by using a well-known property of the Pontryagin square (see Eq.~\eqref{eq:psquare_identity}),
\begin{equation} \label{eq:pontryagin_identity} 
\mathcal P(\alpha + \beta) = \mathcal P(\alpha) + \mathcal P(\beta) + 2\alpha \cup \beta,
\end{equation} 
where $\alpha, \beta \in H^p(X,\ZZ_k)$ and the expression is valid at the level of $\ZZ_{2k}$ cohomology. In the present case $\alpha = k\, n$ is trivial in $H^2(\ZZ_k)$, which implies 
\begin{equation}\label{eq:SPT} 
S_{\text{SPT}}(B) = \frac{2\pi i}{2k}\sum_h \mathcal P(B)\,. 
\end{equation}
This is the SPT action coupled to a background field for the $\ZZ_k$ 1-form symmetry. Note that on our closed periodic lattice, the above action evaluated to a $\ZZ_k$ phase as expected for a spin manifold. 

Now suppose we are on a lattice with boundary where the genuine $\ZZ_{2k}$ nature of the SPT phase appears. The SPT action is no longer background gauge-invariant. Instead (working mod $2\pi i$), 
\begin{align}
&S_{\text{SPT}}(B + dV +k L) - S_{\text{SPT}}(B)=  \\
& \sum_h \frac{2\pi i}{2k} \Big( B \cup dV + dV \cup B + dV\cup_1 dB +dV\cup dV \Big) \nonumber \\
&+ i\pi \Big((B+dV)\cup L + L \cup (B+dV)+ (B+dV)\cup_1 dL\Big)\,. \nonumber
\end{align}
Now using the Leibniz rule and working mod $2\pi i$, this becomes 
\begin{align}
&\sum_h \frac{2\pi i}{2k}\Big( d(B\cup V + V \cup B + V \cup dV) \\
&\quad\quad\quad\quad -dB\cup V- V\cup dB +dV\cup_1 dB \Big)  \nonumber \\
&+i\pi \Big (L \cup (B+dV)-(B+dV)\cup L  + (B+dV)\cup_1 dL \Big)\,. \nonumber
\end{align} 
Again using the cup product identities in Eq.~\eqref{eq:cupidentity} and working mod $2\pi i$ this reduces to 
\begin{multline}
S_{\text{SPT}}(B + dV +k L) - S_{\text{SPT}}(B)=  \\
= \sum_h d\Bigg[\frac{2\pi i}{2k} \left( B\cup V + V \cup B + V \cup dV \right) \\
 + i \pi \left( (B+dV)\cup_1 L + V \cup_1 \frac{1}{k}dB \right)\Bigg]\,,
\end{multline}
which exactly cancels the $\ZZ_{2k}$-valued anomaly in Eq.~\eqref{eq:anomaly}. Therefore, we have established anomaly inflow for the 't Hooft anomaly of the $\ZZ_k$ 1-form symmetry in our lattice CS theory.

%%%%%%%%%%%%%
%{\bf Conclusions.} 
%%%%%%%%%%%%%
\section{Conclusions and outlook}

We have presented a fully regularized Euclidean lattice formulation of compact, $U(1)_k$ Chern-Simons theory with $k$ even. Using this construction, we explored familiar (but subtle) aspects of CS theory such as level quantization, the need for framing, the electric charge of monopoles, and the 't Hooft anomaly for the 1-form symmetry, all at finite lattice spacing. This work provides yet another example which challenges the common lore that certain aspects of continuum quantum field theory cannot be captured on the lattice, and has many worthwhile generalizations and extensions. 

Although we presented our construction on the cubic lattice, all of the features explored in this paper (including the lattice action \eqref{eq:action}) carry over almost verbatim to a general triangulation. On a triangulation, the definitions of (higher) cup products and the framing of Wilson lines depend sensitively on the choice of branching structure (ordering of vertices), making certain aspects more technically involved, but straightforward. 

We focused on the even level case which has an intrinsically three-dimensional definition. The odd level case is more subtle due to the theory being a spin-TQFT. In the `simplest' case of $k=1$, the Wilson line is a fermion, whose spin can be computed via self-linking. However, this non-trivial topological spin cannot be computed using the 't Hooft anomaly for the 1-form symmetry, as there is no 1-form symmetry when $k=1$. A proper lattice formulation of $U(1)_k$ at odd level on the cubic lattice will have to explicitly involve the spin structure, presumably requiring an appropriate definition of $\omega_2$, the second Stiefel-Whitney class. 

Our pure-CS theory can be extended in various ways, for instance by including a Maxwell term or charged matter. The zero modes which required us to study only framed Wilson loops gets lifted by a Maxwell term, and we expect that the long-distance correlation functions of appropriately-defined unframed Wilson loops should match the the correlation functions of untwisted, framed Wilson loops in the pure CS theory.\footnote{Maxwell-Chern-Simons theory also has topological Gukov-Witten operators which generate the $\ZZ_k$ 1-form symmetry. On the lattice, these are ribbons with correlation functions and topological spins determined by the 't Hooft anomaly Eq.~\eqref{eq:anomaly}. }

The main technical ingredients of our lattice formulation are the use of Lagrange multipliers in the modified Villain approach and the (higher) cup products on the cubic lattice. We expect that these tools can be applied to other interesting topological terms in various dimensions which have no obvious definitions on the lattice. This includes the 3d and 4d `Maxwell-Goldstone' models and 4d axion-Maxwell theory, all of which are theories with cubic topological terms and higher-group symmetries~\cite{Cordova:2018cvg,Tanizaki:2019rbk,Brennan:2020ehu,Hidaka:2020izy,Damia:2022rxw}. It would be interesting to try to give rigorous definitions of these theories on the lattice while keeping all global properties intact. Finally, our construction generalizes straightforwardly to torus gauge groups with multiple $U(1)$ factors. It is less obvious how to extend our analysis to non-abelian groups and connect our approach to existing proposals for non-abelian CS terms on the lattice~\cite{Seiberg:1984id}. 

An obvious application of our CS action and its generalizations is to establish exact dualities on the lattice. Of course, it has long been known that particle-vortex duality is exact on the lattice~\cite{Peskin:1977kp,Dasgupta:1981zz}, and more recently it was shown that lattice models in the modified Villain formulation exhibit similar exact dualities~\cite{Sulejmanpasic:2019ytl,Anosova:2022cjm,Gorantla:2021svj}. Dualities between CS-matter theories can in principle be established on the Euclidean lattice simply by comparing worldline representations. For related recent work in the context of fermionic spin models, see~\cite{Chen:2017fvr,Chen:2018nog,Chen:2019wlx,Chen:2021ppt}.

An interesting question which we have not explored here is how to understand the gravitational anomaly of CS theory. It would be interesting to see whether or how the subtle interplay of CS theory with gravity~\cite{Witten:1988hf} manifests itself in our construction.\footnote{We thank Shu-Heng Shao, Nathan Seiberg, and Yuya Tanizaki for raising these points.} In particular CS theory in the continuum, while naively metric-independent, requires the metric in order to gauge fix. However, the dependence on the metric is relatively mild, appearing as a phase of the partition function which depends on the framing of the manifold. In our construction, gauge fixing is not really an issue,\footnote{There needs to be some partial discrete gauge fixing to bring the link gauge fields into a finite interval as is customary in the Villain formulation, but the gauge need not be fully fixed.} as the lattice gauge theory is compact. However the extra zero modes will potentially cause problems, at least on infinite lattices. It would be interesting to see whether changing the choice of cup product leads to the phase ambiguity related to the framing of the manifold expected in the continuum. To understand this, one would have to compute the partition function of our lattice CS theory. This is bound to be subtle because of the staggered symmetry which leads to extra zero modes in the Gaussian operator which must be appropriately modded out. One way this can be done is by introducing a Maxwell term, which would lift the zero modes, and subsequently taking the subtle limit of infinite gauge coupling. 

Finally, another avenue is to formulate compact CS theory on the lattice in the canonical formalism using the Villain Hamiltonian approach~\cite{Fazza:2022fss}. A natural starting point is the modified Villain generalization of the lattice action studied by Eliezer and Semenoff, which is free of zero modes when time is continuous.  Similarly, one should be able to construct the 4d $\theta$-term\footnote{In the Hamiltonian formulation the $\theta$-angle periodicity is only true up to the action by an operator containing the Chern-Simons term.} in the Hamiltonian formulation of the 4d gauge theory~\cite{Fazza:2022fss}.  We leave this for future work.

\acknowledgements

We thank Jing-Yuan Chen, Aleksey Cherman, I\~naki Garcia-Etxebarria, Nabil Iqbal, Zohar Komargodski, Max Metlitski, Shu-Heng Shao, Nathan Seiberg, Yuya Tanizaki, and Mithat \"Unsal for useful comments on the draft of this paper. T. Jacobson would like to thank Durham University and the University of Washington for their generous hospitality during various stages of this work, as well as the organizers of the program ``Topological Phases of Matter: from Low to High energy" at the Institute for Nuclear Theory, where this work was presented. T. Jacobson acknowledges support from the University of Minnesota Doctoral Dissertation Fellowship. T. Sulejmanpasic is supported by the Royal Society University Research Fellowship and in part by the STFC consolidated grant  ST/T000708/1. 

\appendix

%%%%%%%%%%%%%
%{\bf Appendix: (Higher) cup products on the cubic lattice.} 
%%%%%%%%%%%%%
\section{(Higher) cup products on the cubic lattice}
\label{sec:products} 

In this appendix we present explicit expressions for the (higher) cup products on the cubic lattice. As discussed in the main text, the standard cup product of a $p$-cochain ($p$-form) and a $q$-cochain ($q$-form) is a $(p+q)$-cochain ($p+q$-form), while the $\cup_i$ product of a $p$-cochain and a $q$-cochain is a $(p+q-i)$-cochain. In this notation $\cup = \cup_0$.

Two crucial properties of the cup product, which we do not prove here, are that it obeys the Leibniz rule:
\begin{equation}
d(\alpha \cup \beta) = d\alpha \cup \beta + (-1)^p \alpha \cup d\beta, 
\end{equation}
and is only supercommutative up to additional terms involving the cup-1 product:
\begin{align} \label{eq:cupidentityA} 
&\alpha \cup \beta - (-1)^{pq}\,  \beta \cup \alpha =  \\
&(-1)^{p+q+1} \Big[ d(\alpha\cup_1\beta)- d\alpha \cup_1 \beta - (-1)^p\, \alpha \cup_1 d\beta \Big]\,, \nonumber
\end{align}
where $\alpha$ and $\beta$ are $p$- and $q$-cochains respectively. The above pattern continues---the $\cup_i$ product supercommutes only up to terms involving the $\cup_{i+1}$ product:
\begin{align} \label{eq:higher_cupidentity} 
&\alpha \cup_i \beta - (-1)^{pq+i}\,  \beta \cup_i \alpha  = \\
&(-1)^{p+q+1+i} \Big[ d(\alpha\cup_{i+1}\beta)- d\alpha \cup_{i+1} \beta - (-1)^p\, \alpha \cup_{i+1} d\beta \Big]\,. \nonumber
\end{align}
Note that the $\cup_i$ product strictly vanishes unless $i \le p, q $. 

A general combinatorial definition of the higher cup product on the hypercubic lattice is given in Ref.~\cite{Chen:2021ppt}.\footnote{See Eqs.~(27) and (28) in Ref.~\cite{Chen:2021ppt}. Note that the convention we choose here corresponds to swapping all $+$ labels to $-$ and visa-versa in all of their formulas.} For completeness and clarity, we present graphical depictions of the (higher) cup products in 1, 2, and 3 dimensions as well as explicit formulas using notation which is standard in lattice gauge theory. We will not give a general proof of the identities Eqs.~\eqref{eq:cupidentityA},\eqref{eq:higher_cupidentity}, but one can verify that they hold for the specific cases provided below. 

Our notation follows that of Ref.~\cite{Sulejmanpasic:2019ytl}. A $p$-cochain (or $p$-form) $\alpha$ is denoted $\alpha_{x,\mu_1\mu_2 \cdots \mu_p}$ with $x$ the `root' site from which the $p$-chain (or $p$-cell) emanates, and the indices run between $1 \le \mu_i \le d$. Below we will always take $\mu_1 < \mu_2 \cdots < \mu_p$, with explicit minus signs to indicate orientation. 

\begin{itemize}
\item Ordinary cup products in 1 and 2 dimensions, depicted in Fig.~\ref{fig:cup_a}:
\begin{subequations}
\begin{align}
&(\alpha^{(0)} \cup \beta^{(1)})_{x,1} = \alpha_x \, \beta_{x,1}\,, \\
&(\beta^{(1)}\cup \alpha^{(0)} )_{x,1} = \beta_{x,1} \, \alpha_{x+\hat{1}}\,, \\
&(\alpha^{(1)}\cup\beta^{(1)})_{x,12} = \alpha_{x,1}\, \beta_{x+\hat{1},2} - \alpha_{x,2}\, \beta_{x+\hat{2},1}\,, \\
&(\alpha^{(0)} \cup \beta^{(2)} )_{x,12} = \alpha_x \, \beta_{x,12}\,, \\
&(\beta^{(2)} \cup \alpha^{(0)} )_{x,12} =  \beta_{x,12} \, \alpha_{x + \hat{1}+\hat{2}}\,.
\end{align}
\end{subequations}

\begin{figure}[h] 
\centering
\includegraphics[width=0.45\textwidth]{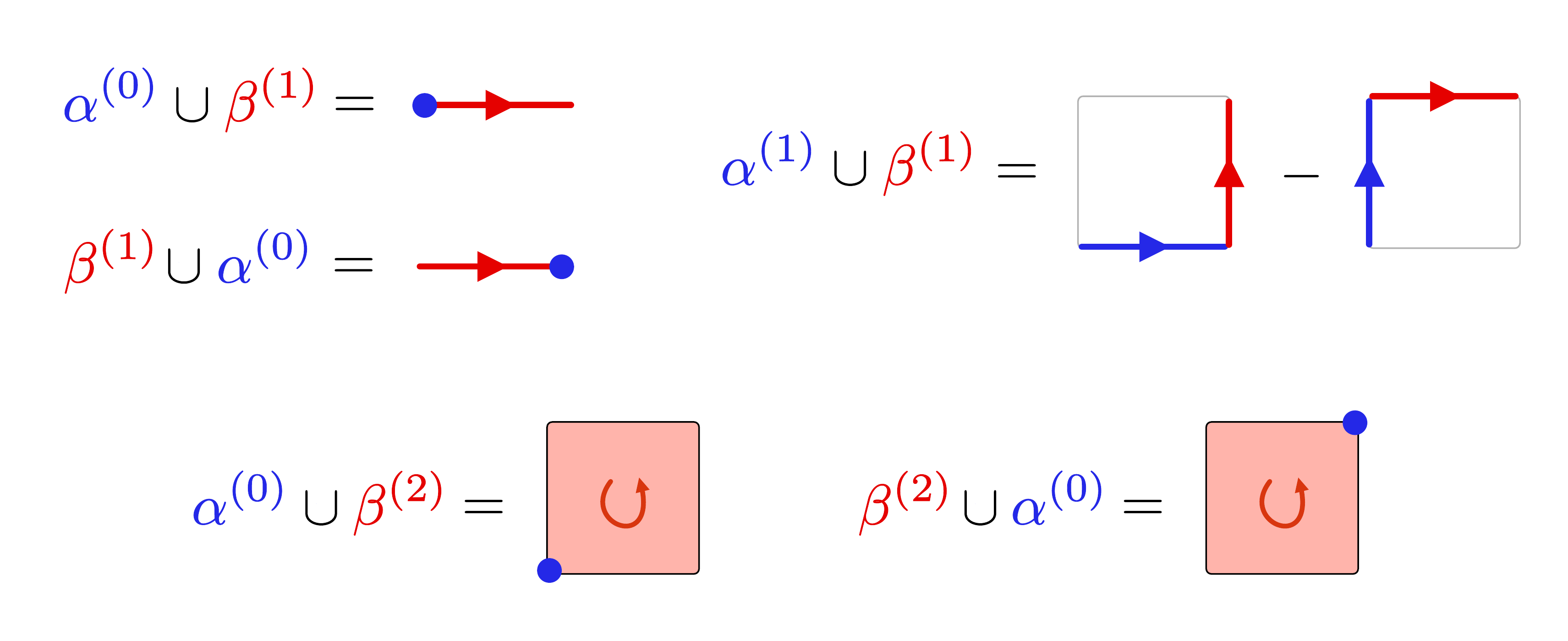}
\caption{Ordinary cup products in 1 and 2 dimensions. } 
\label{fig:cup_a}
\end{figure}

\item Ordinary cup products in 3 dimensions, depicted in Fig.~\ref{fig:cup_b}:
\begin{subequations}
\begin{align}
&(\alpha^{(1)} \cup \beta^{(2)})_{x,123} =  \\
&\quad \alpha_{x,1}\, \beta_{x+\hat{1},23} -\alpha_{x,2}\, \beta_{x + \hat{2}, 13} + \alpha_{x,3}\, \beta_{x + \hat{3},12}\,, \nonumber \\
&(\beta^{(2)} \cup \alpha^{(1)})_{x,123} = \\
&\quad  \beta_{x,23}\, \alpha_{x+\hat{2}+\hat{3},1} -  \beta_{x,13}\, \alpha_{x+\hat{1}+\hat{3},2} +  \beta_{x,12}\, \alpha_{x+\hat{1}+\hat{2},3}\,, \nonumber \\
&(\alpha^{(0)} \cup \beta^{(3)})_{x,123} = \alpha_{x} \, \beta_{x,123}\,, \\
&(\beta^{(3)} \cup \alpha^{(0)})_{x,123} = \beta_{x,123}\, \alpha_{x + \hat{1}+\hat{2} + \hat{3}}\,.
\end{align}
\end{subequations}

\begin{figure}[h] 
\centering
\includegraphics[width=0.45\textwidth]{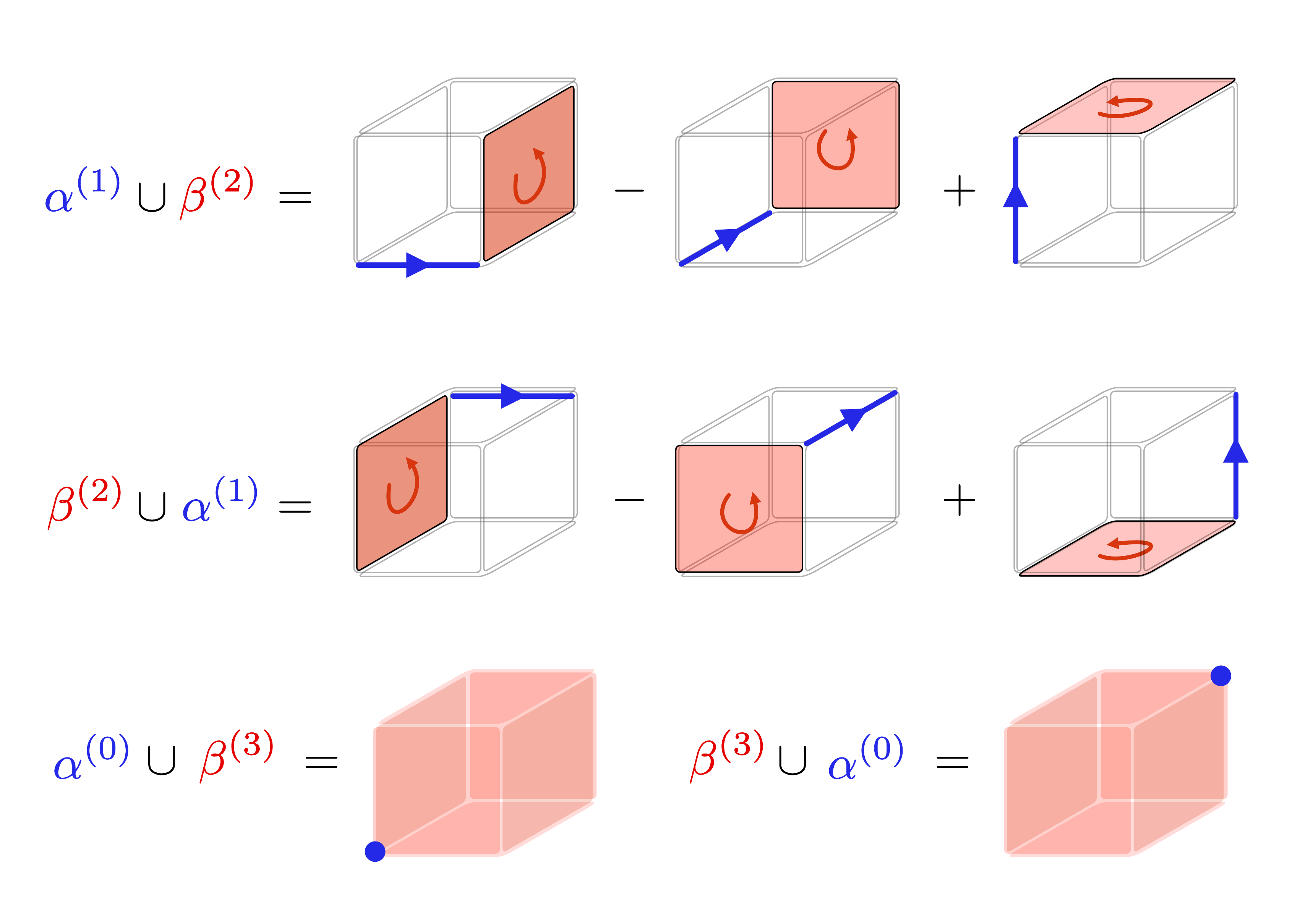}
\caption{Ordinary cup products in 3 dimensions. } 
\label{fig:cup_b}
\end{figure}

\item Higher cup products in 1 and 2 dimensions, depicted in Fig.~\ref{fig:cup_c}:
\begin{subequations}
\begin{align}
(\alpha^{(1)} \cup_1 \beta^{(1)})_{x,1} = \alpha_{x,1} \, \beta_{x,1}\,, \\
(\alpha^{(1)} \cup_1 \beta^{(2)})_{x,12} = -(\alpha_{x,2}+\alpha_{x+\hat{2},1})  \beta_{x,12}\,, \\
(\beta^{(2)} \cup_1 \alpha^{(1)})_{x,12} = \beta_{x,12}(\alpha_{x,1} + \alpha_{x+\hat{1},2})\,, \\
(\alpha^{(2)} \cup_2 \beta^{(2)})_{x,12} = \alpha_{x,12} \, \beta_{x,12}\,.
\end{align}
\end{subequations}

\begin{figure}[h] 
\centering
\includegraphics[width=0.45\textwidth]{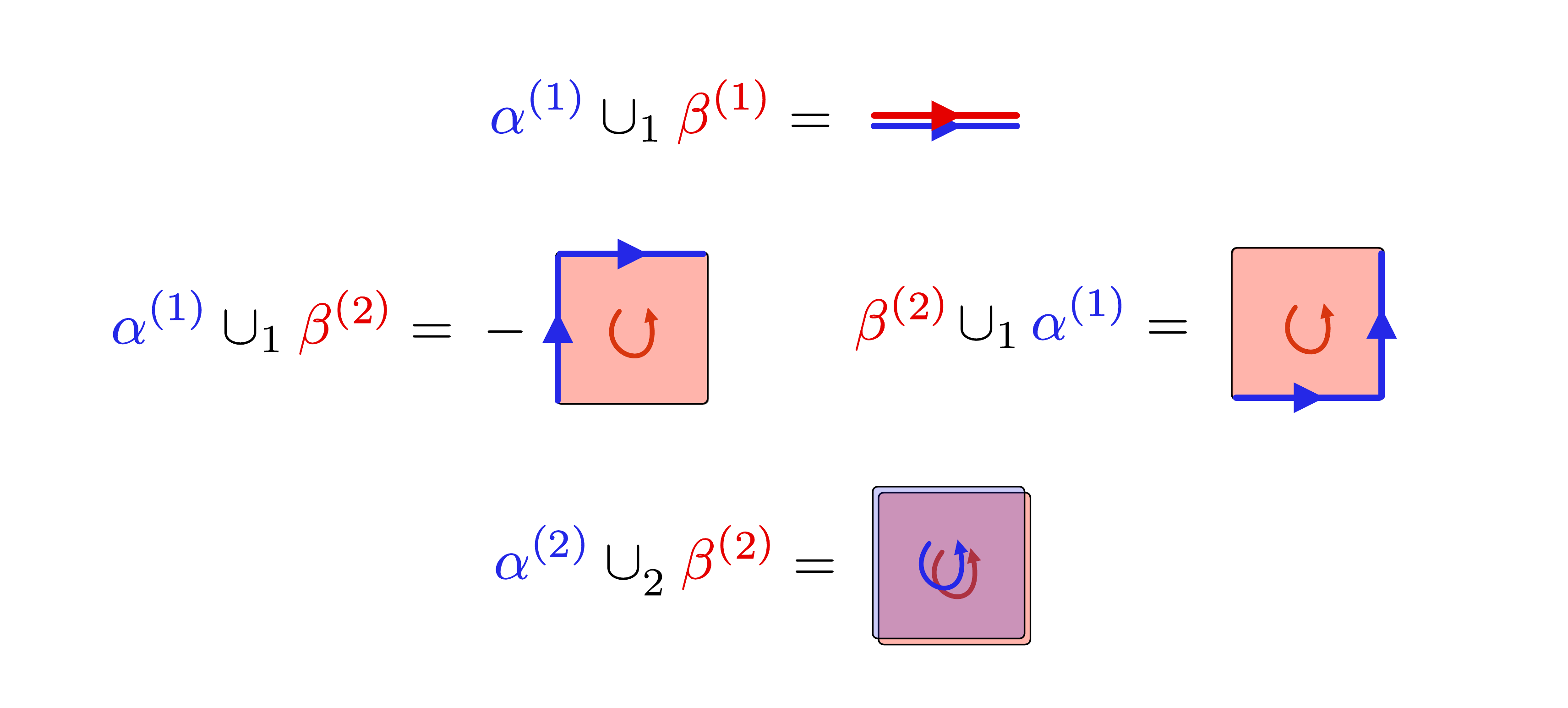}
\caption{Higher cup products in 1 and 2 dimensions. } 
\label{fig:cup_c}
\end{figure}

\item Cup-1 products in 3 dimensions, depicted in Fig.~\ref{fig:cup_d}:
\begin{subequations}
\begin{align}
&(\alpha^{(1)} \cup_1 \beta^{(3)})_{x,123} \\
&\quad= (\alpha_{x,3}+\alpha_{x+\hat{3},2} + \alpha_{x+\hat{3}+\hat{2},1}) \beta_{x,123}\,, \nonumber \\
&(\beta^{(3)} \cup_1 \alpha^{(1)})_{x,123} \\
&\quad = \beta_{x,123}(\alpha_{x,1}+\alpha_{x+\hat{1},2} + \alpha_{x+\hat{1}+\hat{2},3})\,, \nonumber \\
&(\alpha^{(2)}\cup_1 \beta^{(2)})_{x,123} = \alpha_{x,23}(\beta_{x,12} + \beta_{x+\hat{2},13})  \\
&\quad\quad+ \alpha_{x+\hat{2},13}\, \beta_{x,12} - \alpha_{x,13}\, \beta_{x+\hat{1},23} \nonumber \\
&\quad\quad- \alpha_{x+\hat{3},12}( \beta_{x,13} + \beta_{x+\hat{1},23})  \,. \nonumber 
\end{align}
\end{subequations}

\begin{figure}[h] 
\centering
\includegraphics[width=0.45\textwidth]{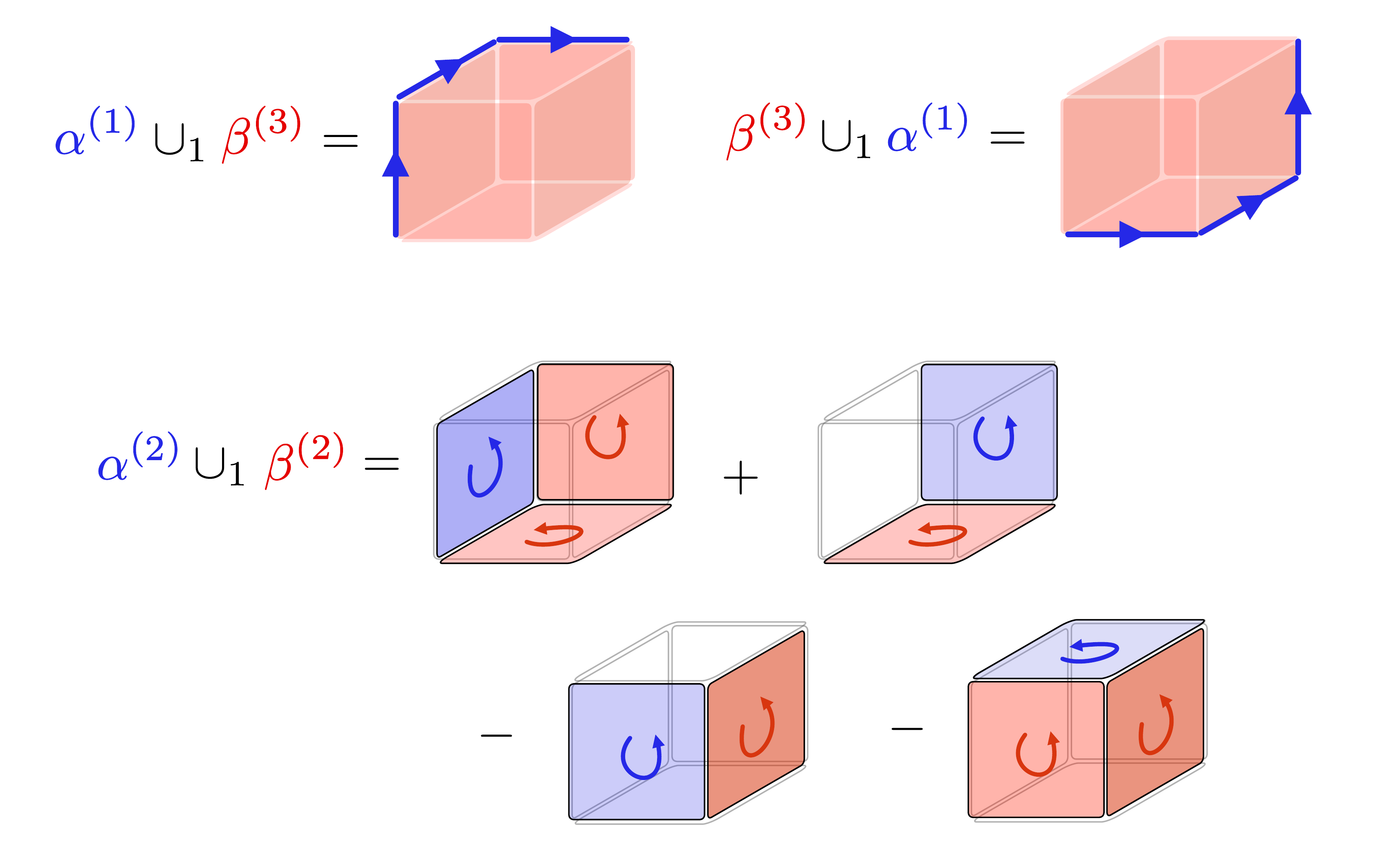}
\caption{$\cup_1$ products in 3 dimensions. } 
\label{fig:cup_d}
\end{figure}

\item Cup-2 and cup-3 products in 3 dimensions, depicted in Fig.~\ref{fig:cup_e}:
\begin{subequations}
\begin{align}
(\alpha^{(2)}\, \cup_2 \, &\beta^{(3)})_{x,123} \\
&= (\alpha_{x,12} + \alpha_{x,23} + \alpha_{x+\hat{2},13})\beta_{x,123}\,, \nonumber \\
(\beta^{(3)}\,  \cup_2 \, &\alpha^{(2)})_{x,123} \\
&= \beta_{x,123}(\alpha_{x,13} + \alpha_{x+\hat{1},23} + \alpha_{x+\hat{3},12})\,, \nonumber \\
(\alpha^{(3)} \,\cup_3\,  & \beta^{(3)})_{x,123} = \alpha_{x,123}\, \beta_{x,123} \,.
\end{align}
\end{subequations}

\begin{figure}[h] 
\centering
\includegraphics[width=0.45\textwidth]{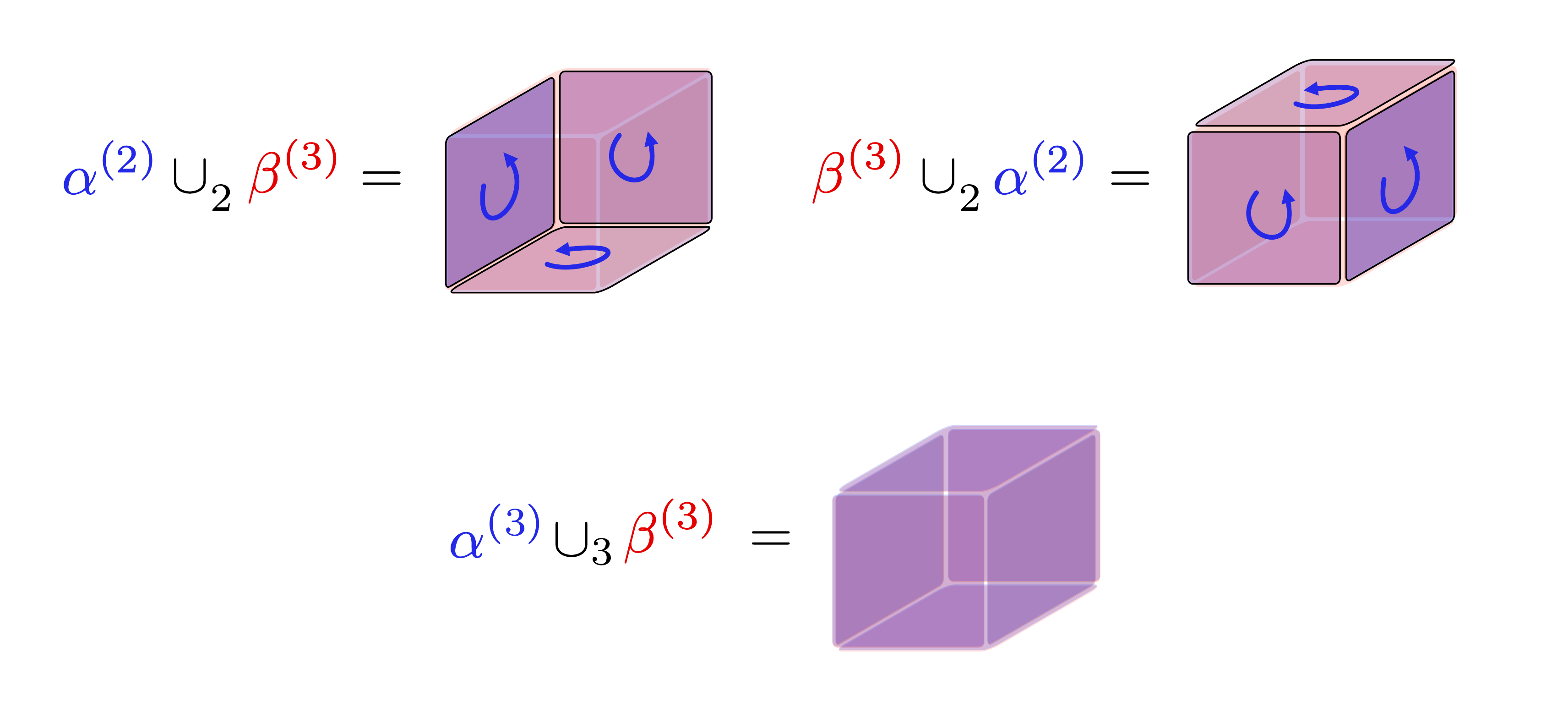}
\caption{$\cup_2$ and $\cup_3$ products in 3 dimensions. } 
\label{fig:cup_e}
\end{figure}

\item Here we only give an explicit formula for the $\cup_1$ product of a $2$-cochain and a $3$-cochain in 4 dimensions needed to define the theta term in Eq.~\eqref{eq:theta} and the Pontryagin square: 
\begin{subequations}
\label{eq:4dcup} 
\begin{align}
&(\alpha^{(2)} \cup_1 \beta^{(3)})_{x,1234} = \\
&\quad(\alpha_{x,34} + \alpha_{x+\hat{3},24} + \alpha_{x+\hat{2}+\hat{3},14} )\beta_{x,123} \\
&+ (\alpha_{x,14} + \alpha_{x+\hat{4},13} + \alpha_{x+\hat{3}+\hat{4},12})\beta_{x+\hat{1},234} \\
&-(\alpha_{x+\hat{4},23} + \alpha_{x+\hat{2}+\hat{4},13})\beta_{x,124}+\alpha_{x,34}\, \beta_{x+\hat{3},124}  \\
&- (\alpha_{x,24} +\alpha_{x+\hat{4},23})\beta_{x+\hat{2},134}+ \alpha_{x+\hat{3}+\hat{4},12}\, \beta_{x,134} \,.
\end{align}
\end{subequations}

\end{itemize}

\section{The anomaly of $U(1)_k$ in the continuum and the Pontryagin square}
\label{sec:pontryagin} 

Consider the $U(1)_k$ Chern-Simons theory in continuum on a 3d Euclidean manifold $M$. Rigorously the theory is defined by using an auxiliary 4d manifold $X$ such that $M=\partial X$, with the following action
\begin{equation}\label{eq:cont_CS}
S=\frac{ik}{4\pi}\int_X F\wedge F 
\end{equation}
where $F=dA$ and where $A$ is a $U(1)$ connection on $X$ which smoothly extends from the connection on $M$. To define the path integral on $M$, one integrates over all gauge fields on $M$ appropriately extended to $X$, with a weight given by $e^{-S}$.\footnote{What is meant by this is that for a particular configuration $A$ on $M$, one picks a 4d manifold $X$ whose boundary is $M$, over which $A$ extends smoothly and then uses \eqref{eq:cont_CS} to compute the weight. Note that it may be necessary to pick a different $X$ for different configurations $A$ on $M$. } For this to make sense, one must make sure that the weight $e^{-S}$ does not depend on the extension of the gauge fields $A$ on $M$ to the gauge fields on $X$. A standard argument shows\footnote{The argument compares two such extensions $X$ and $X'$ and looks at the difference of weights defined via $X$ and $X'$, i.e. $e^{-S_X+S_{X'}}=e^{-\frac{ik}{4\pi}\int_{X\cup (-X')} F\wedge F}=1$ if $k\in \ZZ$ in the spin case and $k \in 2\ZZ$ in the non-spin case. This follows because $\frac{1}{4\pi}\int F\wedge F\in 2\pi\mathbb Z$ on any closed spin $4$-manifold, but can be half-integral on a non-spin manifold.} that this is true for any $k\in \mathbb Z$ on a spin manifold, and is true only for even $k$ on a non-spin manifold.

The CS theory has $\mathbb Z_k$ 1-form symmetry, for which we can turn on background fields $\tilde B$ by replacing $F\rightarrow F+\tilde B$, with $\tilde B$ the $2$-form $\mathbb Z_{k}$ gauge field,\footnote{For the purpose of the continuum description we simply set $k \tilde B=dY$, where $Y$ is a properly quantized 1-form $U(1)$ gauge field.} i.e. $e^{i \int_\Sigma \tilde B}$ is a $\mathbb Z_{k}$ phase. One way to characterize the 't Hooft anomaly for this symmetry is the failure of the CS action coupled to background fields to be independent of the extension to $X$, i.e. by the integral
\begin{multline}
\frac{ik}{4\pi}\int (F+\tilde B)\wedge (F+\tilde B)=\\=\frac{ik}{4\pi}\int F\wedge F+\frac{ik}{2\pi}\int F\wedge \tilde B+\frac{ik}{4\pi}\int \tilde B\wedge \tilde B\;.
\end{multline}
It is easy to convince oneself that the second term is $0\bmod 2\pi$ on a closed manifold, but the third one is in general not. We want to understand the degree of the anomaly, i.e. in what group the phase in the last term take values in. It is useful to consider the normalization $B=\frac{1}{2\pi}k\tilde B$, so that $\int B\in \mathbb Z$. Consider therefore
\begin{equation}
e^{\frac{i k}{4\pi}\int \tilde B\wedge \tilde B}=e^{\frac{i2\pi}{2k}\int B\wedge B}\;.
\end{equation}
Now we must distinguish between spin and non-spin manifolds. Firstly we start with a spin manifold, in which case $\int B\wedge B$ is always an even integer on a closed manifold, and so the above phase is a $\mathbb Z_k$ phase. On a more general non-spin manifold $\int B\wedge B$ can take any integer value in general, and the phase is $\mathbb Z_{2k}$. However, recall that for odd $k$, the CS theory is not well-defined on a non-spin manifold, and so we conclude that the anomaly in general has degree $2k$ for even $k$ and degree $k$ for odd $k$. 

In fact this is a direct reflection of the properties of the Pontryagin square. We will now briefly describe the correspondence between the Pontryagin square in the continuum and on the lattice. Much of this discussion can be found in one form or another in Refs.~\cite{Kapustin:2013qsa,Kapustin:2014gua}. A $\ZZ_k$ gauge field $B$ is a member of cohomology $H^p(M,\ZZ_k)$, where $M$ is the spacetime manifold and $p$ is the form degree of $B$. We will take $p$ to be even in what follows.

In the continuum we can describe $B$ by a representative of De Rham cohomology, i.e. it is a flat $p$-form with $\int_{\Sigma_p} B\in \mathbb Z$ where $\Sigma_p$ is any $p$-cycle of the manifold. Then we can construct a wedge product 
\begin{equation}\label{eq:BwB}
B\wedge B\;.
\end{equation}
But we actually want to think of $B$ as a member of $H^p(M,\mathbb Z_{k})$ and not $H^p(M,\mathbb Z)$. To achieve that, we impose gauge invariance under $B\rightarrow B+k C$ where $C\in H^p(M,\mathbb Z)$, so that $B$ can be thought of as a $\mathbb Z_k$ gauge field, i.e. $e^{i\frac{2\pi s}{k}\int B}$ are well defined for integer $s$ only. Then the statement is that on any manifold \eqref{eq:BwB} is well-defined $\bmod\ 2k$ if $k$ is even and $\bmod \ k$ if $k$ is odd. In other words $B\wedge B\in H^{2p}(M,\mathbb Z_{2k})$ for $k$ even and $B\wedge B\in H^{2p}(M,\mathbb Z_{k})$ for $k$ odd. To see this, we note that under the transformation $B\rightarrow B+kC$, Eq.~\eqref{eq:BwB} transforms as
\begin{equation}
B\wedge B\rightarrow B\wedge B+2k B\wedge C+k^2C\wedge C\;.
\end{equation}
To find the cohomology group for which the above transformation is invisible, we integrate both sides on an arbitrary manifold, and find
\begin{equation}
\int B\wedge B \to \int B\wedge B+\bmod \begin{cases}
2k& \text{for $k$-even}\\
k & \text{for $k$-odd}\;,
\end{cases}
\end{equation}
which establishes the result that for even $k$, $B\wedge B\in H^{2p}(M,\mathbb Z_{2k})$ and for odd $k$, $B\wedge B\in H^{2p}(M,\mathbb Z_{k})$. Notice that the crucial property to establish the result for even $k$ was the commutativity of the cup product $C\wedge B=B\wedge C$.

Now let us return to the anomalous phase, given by
\begin{equation}
e^{i\frac{2\pi}{2k}\int B\wedge B}\;.
\end{equation}
Consider first the even $k$, so that the phase well defined and is a $\mathbb Z_{2k}$ phase on a general (potentially non-spin) manifold, as it should be. On the other hand if $k$ is odd, the above expression is only well-defined if the manifold is spin, in which case the phase lies in $\mathbb Z_k$. So the anomaly is described by the Pontryagin square of $B$.

On the lattice we can work directly at the level of $\ZZ_k$ cohomology, and take $B \in H^2(M,\ZZ_k)$ (i.e. $dB = 0 \text{ mod }k$). Consistency requires invariance under $B \to B + dV + kL$  with $V \in C^1(M,\ZZ)$ and $L \in C^2(M,\ZZ)$. We now start with the analog of the wedge product Eq.~\eqref{eq:BwB}, 
\begin{equation}
B \cup B,
\end{equation}
and ask whether this is a well-defined product at the cohomology level. Unlike in the continuum, where the $\wedge$ product in de Rham cohomology is supercommutative, the $\cup$ product at the cochain level is not. Let us consider a replacement $B\rightarrow B+C$, where we will set $C=dV$ and $C=kL$ at the end, to check the gauge transformation. We have that
\begin{multline}
B\cup B\rightarrow B\cup B+C\cup B+B\cup C+C\cup C=\\= B\cup B+2B\cup C+C\cup C\\-d(C\cup_1 B)+C\cup_1 dB+dC\cup_1 B 
\end{multline}
where we used \eqref{eq:cupidentity}. The second line is very much like the one in the continuum, and if we set $C=dV$ or $C=kL$ it is easily verified that it reduces to $B\cup B\bmod 2k$ for even $k$ and $B\cup B \bmod k$ for odd $k$. The first term of the third line is a total derivative and vanishes after the sum over the appropriate $2p$-cells. The second and third terms in the third line vanish mod $k$. 

When $k$ is even we can improve this product to get something which is well-defined mod $2k$. To do this we must cancel the additional terms above, i.e. introduce a counter-term to $B\cup B$ such that the combination transforms by terms which vanish mod $2k$. Such a term must be bilinear in $B$, and it should involve the $\cup_1$ product. There are only two such terms we can write:
\begin{equation}
B\cup_1 dB \text{ or } dB\cup_1 B\;.
\end{equation}
However these two terms are completely equivalent mod $k^2$ and therefore we can use either.\footnote{Remember that we are trying to construct a class of degree $k$ for odd $k$ and $2k$ for even $k$, which in both cases are divisors of $k^2$.} Moreover, $B\cup_1 dB=-B\cup_1dB\bmod 2k$ so even the sign is irrelevant. Hence we land on 
\begin{equation}
\mathcal P(B)=B\cup B+B\cup_1dB\,.
\end{equation}  
Let us check if this is indeed well-defined under the transformation $B\rightarrow B+C$ with $C$ either $dV$ or $C=kL$. We have that
\begin{multline}
\mathcal P(B)\to \mathcal P(B)+2C\cup B+C\cup C\\-d(C\cup_1 B)+2C\cup_1 dB\\
+dC\cup_1 B+B\cup_1 dC\,.
\end{multline}
The first two lines are not problematic. Finally we have the term $dC\cup_1 B+B\cup_1 dC$, which is identically zero if $C=dV$, but one must check what happens if $C=kL$. Now using the identity for the commutation of the higher cup product~\eqref{eq:cupidentityA}, so that we have
%\footnote{The general formula for $\cup_i$ product is 
%\begin{multline}
%d(\alpha^{(p)}\cup_i\beta^{(q)})=d\alpha^{(p)}\cup_{i}\beta^{(q)}+(-1)^p\alpha^{(p)}\cup_{i}d\beta^{(q)}\\+(-1)^{p+q-i}\alpha^{(p)}\cup_{i-1}\beta^{(q)}+(-1)^{pq+p+q}\beta^{(q)}\cup_{i-1}\alpha^{(p)}
%\end{multline}
%where $\alpha^{(p)}$ is a $p$-cochain and $\beta^{(q)}$ is a $q$ cochain.},  
\begin{equation}
dC\cup_1 B+B\cup_1 dC= - d(dC\cup_2B)-dC\cup_2 dB
\end{equation}
Now setting $C=kL$ we have that $dC\cup_2dB=k\, dL\cup_2dB=0\bmod k^2$ and so $\mathcal P(B)$ is well defined mod $2k$ (resp. $k$) for $k$ even (resp. odd) as expected.

Finally, let us verify the identity Eq.~\eqref{eq:pontryagin_identity}. Let $\alpha, \beta \in H^2(\ZZ_k)$. Then
\begin{align} \label{eq:psquare_identity} 
 &\mathcal P(\alpha + \beta) = \mathcal P(\alpha) + \mathcal P(\beta) + \alpha \cup \beta + \beta \cup \alpha \\
&\hspace{3.5cm}+ \alpha \cup_1 d\beta + \beta \cup_1 d\alpha \nonumber \\
&= \mathcal P(\alpha) + \mathcal P(\beta) +2 \alpha \cup \beta  \nonumber \\
&\quad- d(\beta \cup_1\alpha) +d\beta \cup_1 \alpha + 2\beta \cup_1 d\alpha + \alpha \cup_1 d\beta  \nonumber \\
&= \mathcal P(\alpha) + \mathcal P(\beta) +2 \alpha \cup \beta  \nonumber \\
&\quad - d(\beta \cup_1\alpha) + 2\beta \cup_1 d\alpha - d(d\beta\cup_2\alpha) - d\beta \cup_2 d\alpha \,.\nonumber
\end{align}
All of the terms in the last line are exact or multiples of $2k$, as desired. 

\newpage

%\ts{I propose we erase below.}
%
%Hence we find a gauge variation
%\begin{subequations}
%\begin{align}
%\frac{2\pi i}{2k}\sum_c ( B \cup dV + dV \cup B + dV \cup dV ) \\
%+ k ((B+dV) \cup L + L \cup (B+dV)) +k^2 L\cup L  \\
%= \frac{2\pi i}{2k}\sum_c (-2\, dB\cup V + dV\cup_1 dB) \\
%+k ( 2\, (B+dV) \cup L + k\, L \cup L ) \\
%+k (dL \cup_1 (B+dV) + L \cup_1 dB)\,,
%\end{align}
%\end{subequations}
%dropping total derivatives. We can drop the $L\cup L$ term if we take $k \in 2\ZZ$. However, the variation mod $2\pi i$ is non-vanishing. Using $dB = 0$ mod $k$, and the fact that $k^2 = 0 \text{ mod } 2k$, we are left with  
%\begin{align}
%&\frac{2\pi i}{2k}\sum_c dV \cup_1 dB + k\, dL\cup_1 (B+dV)  \\
%&= \frac{2\pi i}{2k}\sum_c dV \cup_1 dB - k\, (B+dV) \cup_1 dL- k\, dL\cup_2 dB  \nonumber \\
%&= - \frac{2\pi i}{2k}\sum_c dV \cup_1 dB + k\, (B+dV) \cup_1 dL\,,
%\end{align}
%where we used the higher-cup product identity $dL \cup_1 B + B\cup_1 dL = -d(dL\cup_2 B) - dL\cup_2 dB$. Now this can be exactly cancelled mod $2\pi i$ by the variation of $B \cup_1 dB$, and we land on the gauge-invariant definition
%\begin{align}
%\mathcal P(B) \equiv B \cup B + B\cup_1 dB\, .
%\end{align}

\bibliographystyle{utphys}
\bibliography{lattice_cs}

\end{document}